# Geometric Interpretation of 3-SAT and Phase Transition.

Frederic Gillet (frederic.gillet@gmail.com)

## Abstract:
Interpretation of 3-SAT as a volume filling problem, and its use to explore the SAT/UNSAT phase transition.

Notes:
- Source code available at *https://github.com/DrMirakle/three_sat.git*
- The explored algorithms are not meant to be practical and/or efficient 3-SAT solvers.
- This geometric interpretation is fundamentally equivalent to using boolean arithmetic.
- This article does not make any new claim regarding $P \stackrel{?}{=} NP$
- At least one statement in this article is wrong (but it could be this one).

## Introduction:

3-SAT is an NP-complete type of problem, where an "instance" consists of:

- A set of $N$ input boolean variables $\{ v_1, v_2, v_3, \dots v_N\}$, taking value true or false. Using those variables we define literals, a literal on variable $v_i$ is either $v_i$ or its negative $\bar{v}_i$.
- A set of $M$ clauses, where each clause is a disjunction (OR, |) of three literals. E.g. $(\bar{v}_1 | v_1 | \bar{v}_4)$.
- A formula $F$, as a conjunction (AND, &) of the $M$ clauses.

An example consists in four variables $\{ v_1, v_2, v_3, v_4\}$ and two clauses with
$F = (v_1|\bar{v}_2|\bar{v}_3)$ & $(v_1|v_2|v_4)$

The problem consists in answering the question "is there a set of values for the input variables that sets $F$ as true?"

In our example the answer is positive, with the set of values $\{ v_1$= true, $v_2$= true, $v_3$= true, $v_4$= true$\}$
As $N, M$ grow, the time cost for solving the problem grows exponentially $O(2^{N+M})$, at least for the hardest cases.

# Solution Space as a Volume

For $N$ variables, the total number of possible solutions is $2^N$. This solution space can be interpreted as a volume in N dimensions, each dimension having two possible values: 0 (false),1 (true), i.e. a binary space.

Figure 1. shows the binary space for $N$ = 3, i.e. a cubic space with $2^3$= 8 cells.

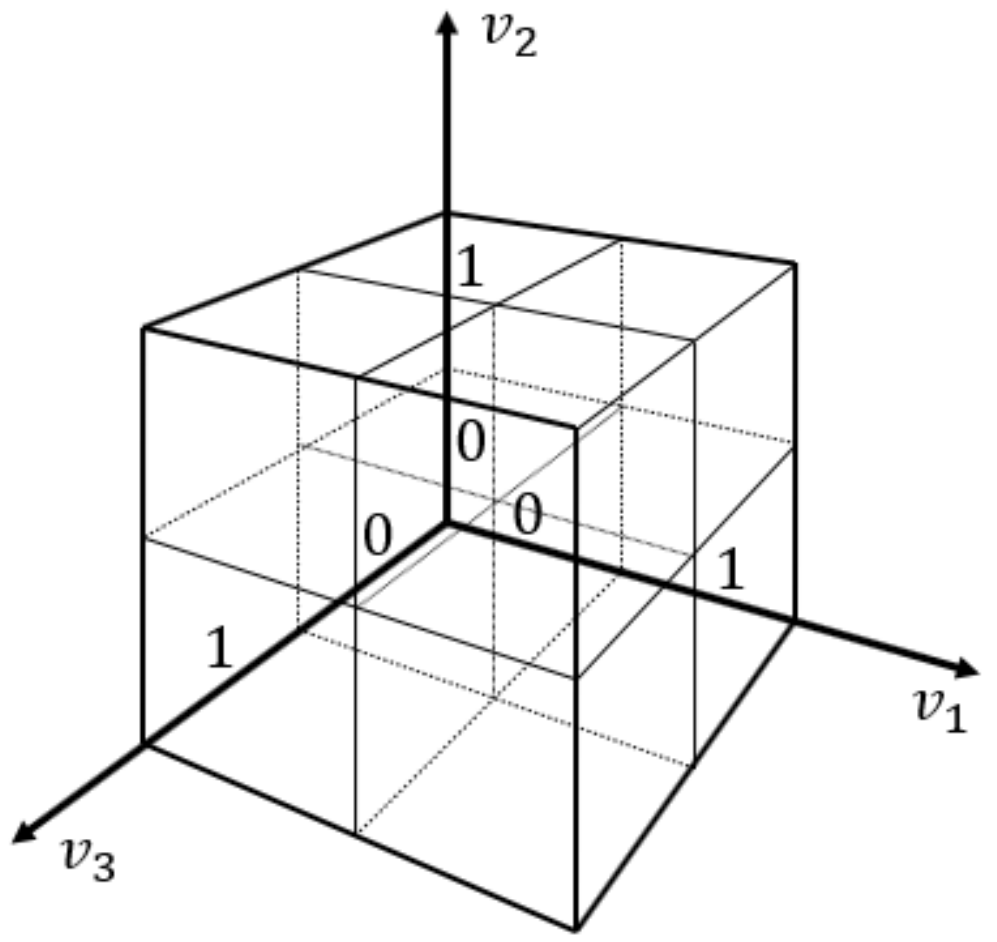

Figure 1 - Solution space with N variables.

The so-called conjunctive normal form of the formula ($N$=6)

$F$ = $(\bar{v}_2 | \ v_3 \ | \ v_5 \ ) \ \& \ (v_1 \ | \ \bar{v}_3 \ | \ v_6 \ ) \ \& \ (v_2 \ | \ v_4 \ | \ \bar{v}_6 \ )$ ...

can be reformulated in an equivalent disjunctive normal form:
$F = \overline{(v_2 \ \& \ \bar{v}_3 \ \& \ \bar{v}_5) | \ (\bar{v}_1 \ \& \ v_3 \ \& \ \bar{v}_6) | (\bar{v}_2 \ \& \ \bar{v}_4 \& \ v_6) \ ...}$

So, each clause is transformed into a more restrictive conjunctive version, and the clauses are now all linked in an additive way.
In particular, if we consider a single clause $(\bar{v}_2 | \ v_3 \ | \ v_5)$, we see that it would only ever be false for the unique triplet $v_2$ = 1, $v_3$ = 0, $v_5$ = 0, regardless of the values of all the other variables, which can be either at 0 or 1, and we represent this with an ‘x’

| $v_1$ | $v_2$ | $v_3$ | $v_4$ | $v_5$ | $v_6$ |
|---|---|---|---|---|---|
| X | 1 | 0 | x | 0 | x |

this expression 'x10x0x' represents a subspace in the binary space of total size $2^6$ where there can be no solutions, i.e. $F$ is guaranteed to be false in that subspace.

Then the next clause $(v_1 \mid \bar{v}_3 \mid v_6)$ defines another volume where there is no solution:

| $v_1$ | $v_2$ | $v_3$ | $v_4$ | $v_5$ | $v_6$ |
|---|---|---|---|---|---|
| 0 | x | 1 | x | x | 0 |

The volumes for those two clauses, 'x10x0x' + '0x1xx0', are additive and represent the subspace where their combination $(\bar{v}_2 \mid v_3 \mid v_5)$ & $(v_1 \mid \bar{v}_3 \mid v_6)$ has no solutions. Therefore, as new clauses are considered/added one by one, the total "no-solution" volume is <u>monotonically increasing</u>, growing steadily until at some point either the entire volume is filled or all the clauses have been added but there is still some empty space left. That empty space represents the set of all solutions to the problem.

The question is how to represent the total volume resulting from the combination of multiple expressions.
For N=3, a simple example is the sum of the expression xx0 and 01x

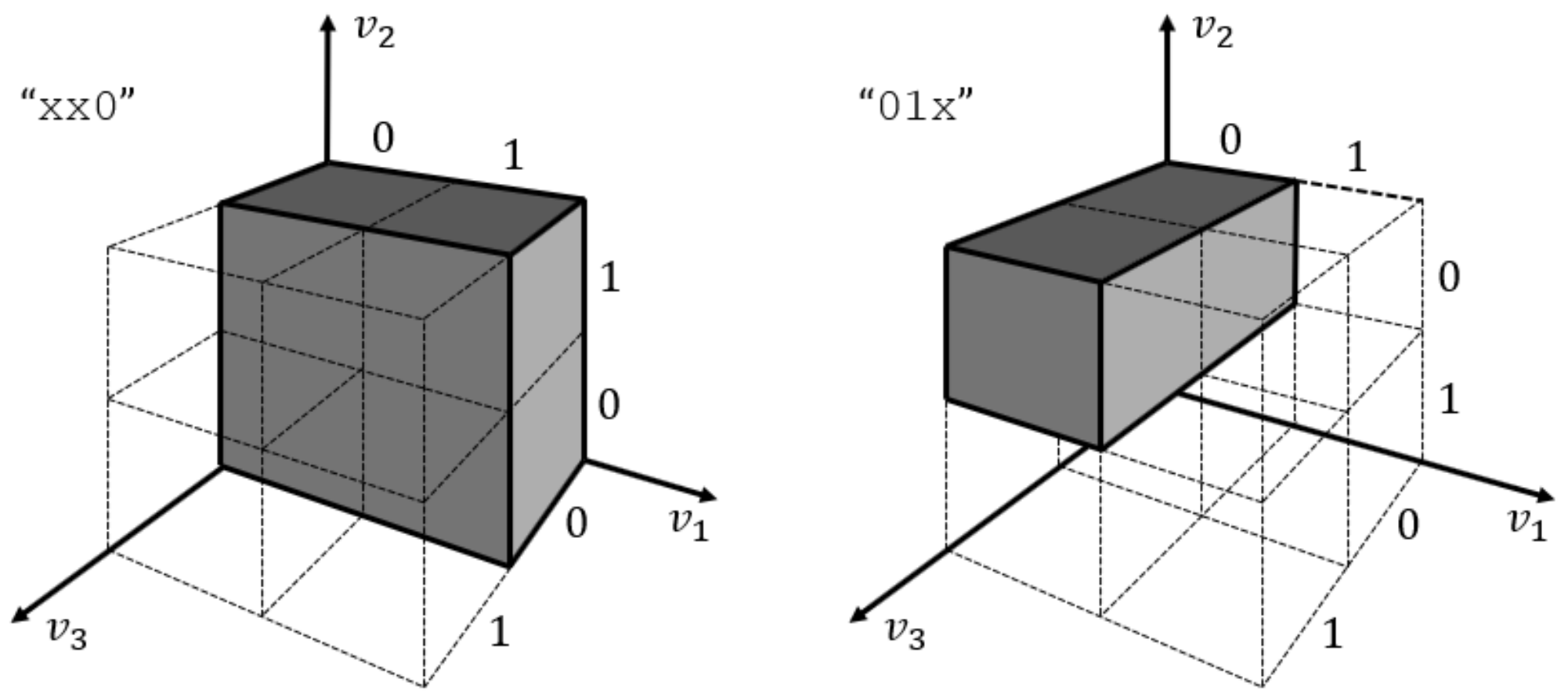

Figure 2 – No-solution volumes for the two expressions.

The resulting combined no-solution volume is shown in Figure 3.

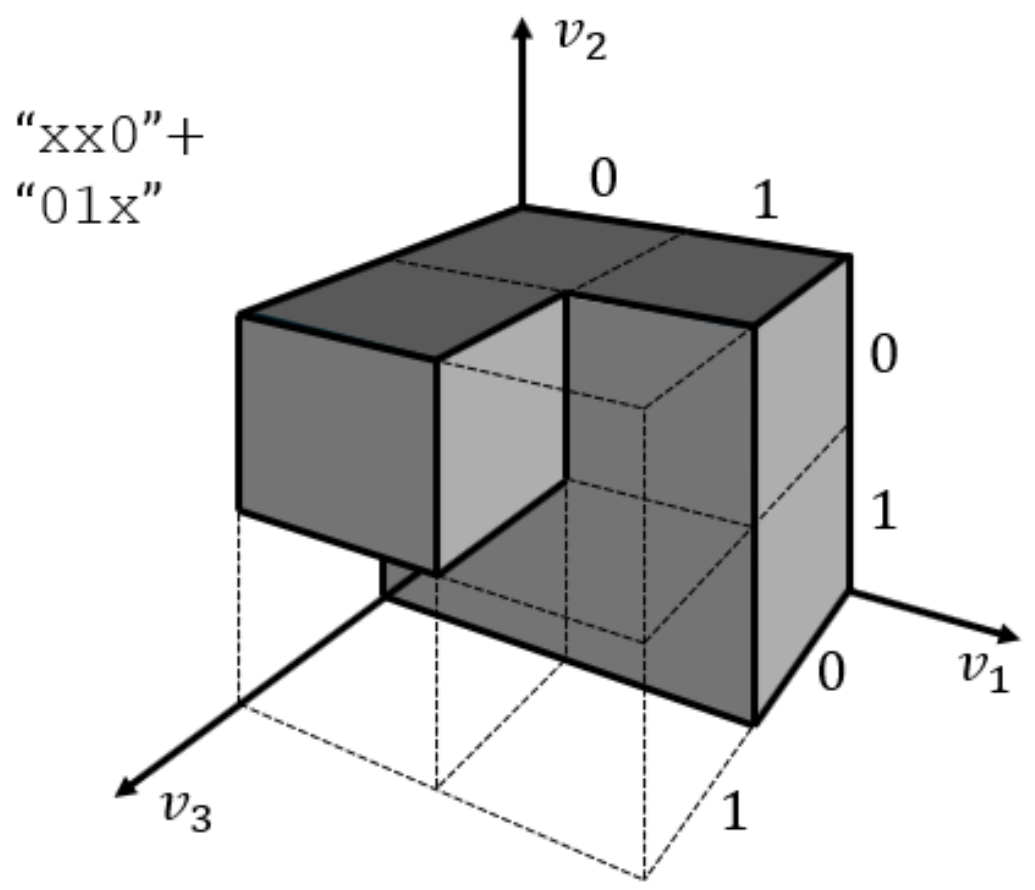

Figure 3 – No-solution volume for the two combined expressions.

A useful representation of the total occupied volume is necessary to be able to extract solutions. Such a representation requires a description in terms of expressions that do not have any overlap.
Two expressions do not overlap (i.e. their volumes have no intersection) when they have at least one common variable that is at 0 in one expression and at 1 in the other expression. In all other cases their volumes overlap.
E.g. the following two expressions do not overlap because the third variable is at 0 and 1:

```
x100xx0
xx10x1x
```

This is because along the dimension axis of that dimension the entire volume is divided in two independent parts.
the following two expressions do overlap

```
000xxxx
xxxx111
```

In order to add up two such expressions, we need to remove the intersection volume from one of the two expressoins, by repeatedly expanding on the dimensions that overlap:

```
000xxxx   000xxxx   000xxxx   000xxxx
xxxx111 → 1xxx111   1xxx111   1xxx111
          0xxx111 → 01xx111   01xx111
                    00xx111 → 001x111
                             (000x111)
```

The expression '`000x111`' can be dropped because it is already entirely included in '`000xxxx`'.
So

```
  000xxxx
+ xxxx111
 _____________
  000xxxx
  1xxx111
  01xx111
  001x111
```

As expected, the resulting 4 expressions have no overlap.
The operation that reverses the expansion described above is a 'merge' – when two expression have no overlap and differ only in a single variable (the one that has opposite values 1/0 in each expression), the two can be merged:

```
  0xx10x
+ 0xx00x
 _____________
  0xxx0x
```

An additional tweak consists in using a sorted binary tree to represent the occupied volume.
This is analog to the "quad tree" structure commonly used in 3D graphics.
The 4 expressions above can then be rearranged by sorting a node at 0 before a corresponding 1. If a node is at '`x`', it will have to be split if a branch with 0 or 1 already exists:

```
000xxxx
001x111
01xx111
1xxx111
```

The notation can be compressed by removing common parent branches up to the point where there is a split, subbranches are separated by the character '|':

```
000xxxx|1x111|1xx111|1xxx111
```

For the simple 3D example of Figure 3, the sum '`xx0 + 01x`' becomes

```
xx0                 xx0              011                 011                 011
01x →"expand"→ 011 →"sort"→  xx0 →"expand"→ 0x0 →"expand"→ 000 →
                   (010)                                 1x0                 010
                                                                             1x0

011              000                000
000 →"sort"→ 010 →"merge"→ 01x → 000|1x|1x0
010              011                1x0
1x0              1x0
```

Note that the final expression '000|1x|1x0' is not more compact (in terms of character count) than the original two non-overlapping expressions 'xx0+011', but the fact that it is in lexical order makes it easier to manipulate.

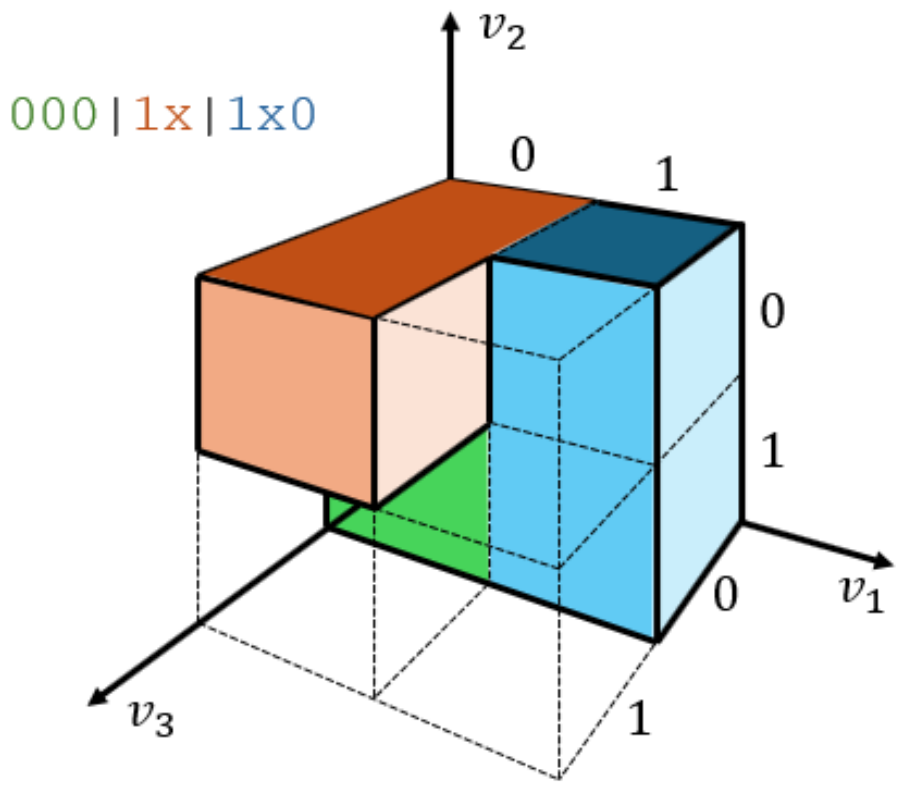

Figure 4 - The total no-solution volume for expression "000|1x|1x0" consisting in three non-overlaping expressions.

The binary tree takes the first variable $v_1$as the root, followed by $v_2, v_3, \ldots v_n$. This order is somewhat arbitrary. Changing the order of the variables can reduce the size of the representation.

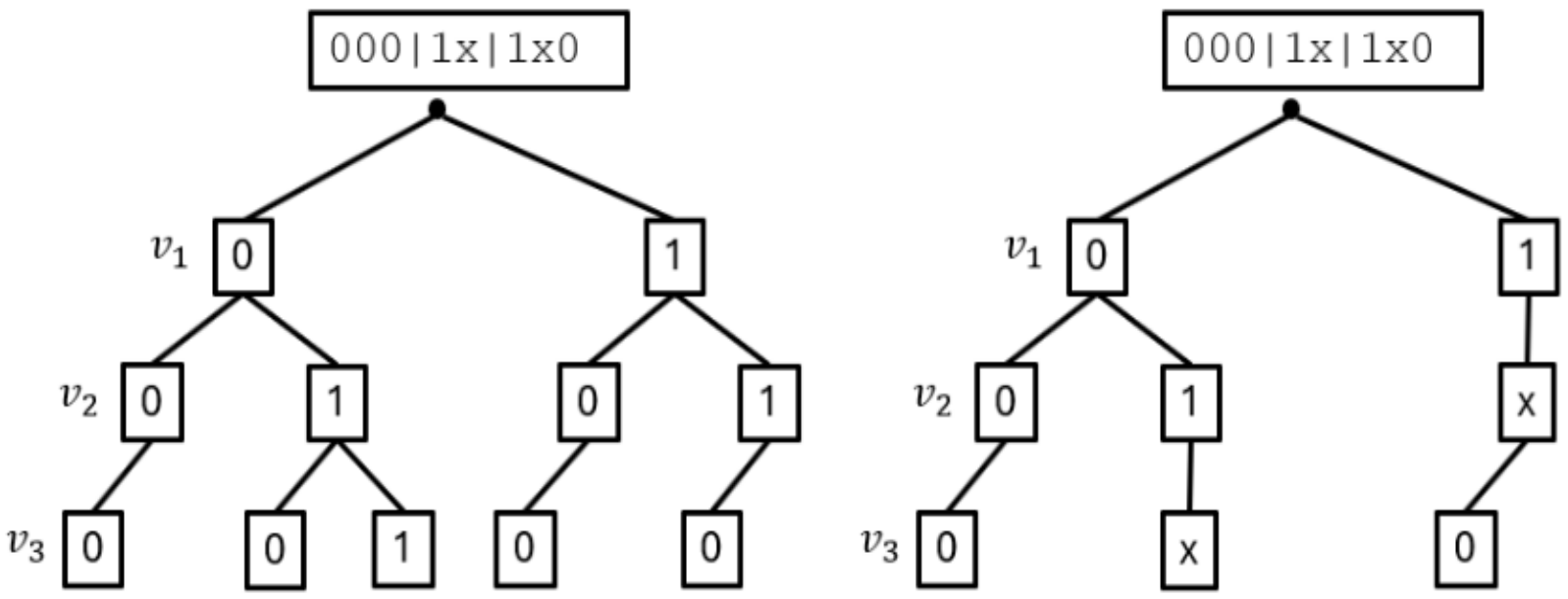

Figure 5 - Binary tree representations of the non-solution volume.

Note that, regardless of N, each triplet clause always covers a subspace which volume is 1/8th (=1/2$^3$) of the total space. If the problem was 4-SAT, each clause would cover a volume that is 1/16th(=1/2$^4$) of the total space.

E.g. for N=6 and a triplet clause $(\bar{v}_2 | \, v_3 \, | \, v_5 \,)$ , the volume of its subspace described by 'x10x0x' is

$$V_c = \; 2 * 1 * 1 * 2 * 1 * 2 = 8 = \; 2^{(6-3)} = 2^{(6-3)} = {2^N}/_8 = {TotalVolume}/_8$$

In other words, the volume of any subspace expression can be computed by counting the occurrences of 'x', $C_X$, and then its volume is $2^{C_X}$, which also gives the number of non-solutions specified by the expression.
The expression 'xxxxxx' for N=6 represents the full volume of the solution space, i.e. there are no solution.
The empty expression '' represent the empty volume, i.e. every cell in the entire space is a solution.

Since each clause is 1/8th of the total volume, it takes at least 8 clauses to induce no solution. 8 such clauses can be trivially built by just creating the 8 combinations of the same chosen 3 variables, with

$$F = (v_1|v_2|v_3)\&(v_1|v_2|\bar{v}_3)\&(v_1|\bar{v}_2|v_3) \ldots (\bar{v}_1|\bar{v}_2|\bar{v}_3)$$

With the volume representations

```
000xxx...
001xxx...
010xxx...
...
111xxx...
```

Each of those subspaces has no overlap with any of the others (because they all differ by at least one variable value that's not in X), so their expressions add up to 'xxxxx...', i.e. there are no solution, as expected. Note that, as we add clauses, once we reach the full volume 'xxxx' at any time, there is no point in processing any more clauses.

## Extracting Solutions

Once the full binary tree describing the non-solutions is built, extracting actual solutions is straightforward.
As an example, here is an instance for N = 4 and M = 10, with the clauses

$\{v_1, v_2, v_3\}$,
$\{\bar{v}_1, v_2, v_3\}$,
$\{v_1, \bar{v}_2, v_3\}$,
$\{\bar{v}_1, v_2, \bar{v}_3\}$,
$\{v_1, \bar{v}_2, \bar{v}_3\}$,
$\{\bar{v}_1, \bar{v}_2, \bar{v}_3\}$,
$\{v_1, v_2, \bar{v}_4\}$,
$\{\bar{v}_1, \bar{v}_2, \bar{v}_4\}$,
$\{\bar{v}_2, v_3, v_4\}$,
$\{v_2, \bar{v}_3, v_4\}$

When clauses are processed in that order, we get the following evolution for the solution space:

```
  ""          ("" is the full empty volume with all 2^4 = 16 solutions)
+ 000x
_________
  000x
+ 100x
_________
  x00x
+ 010x
______________
  0x0x|100x
+ 101x
______________
  0x0x|10xx
+ 011x
___________________
  000x|1xx|10xx
+ 111x
________________________
  000x|1xx|10xx|11x
+ 00x1
____________________________
  000x|11|1xx|10xx|11x
+ 11x1
________________________________
```

```
  000x|11|1xx|10xx|101|1x
+ x100
______________________
  000X|11|1xx|1xxx
+ x010
__________
  xxxx
```

the entire solution space is filled, there are no solution

When all 10 clauses are considered, there is no solution.
It is easy to generate an instance with only one (or few) solution by dropping the last clause that makes the tree full.
In this example, if we drop the last clause $\{v_2, \bar{v}_3, v_4\}$, we get the tree

```
000x|11|1xx|1xxx
```

We can expand and count the volume of each branch to find its solution count:

```
    000x → 1*1*1*2  = 2
    0011 → 1*1*1*1  = 1
    01xx → 1*1*2*2  = 4
    1xxx → 1*2*2*2  = 8
______________________________
```

occupied volume = 15
solution count = 2^4 – 15 = 1

There is only a single solution, which can be extracted (Figure 6) by finding a path from root to leaf that doesn't included any 'x'. The variable values for the solution are the path values, except for the leaf, which is the opposite.

This solution `0010`, which makes the 9 first clauses true, and the last one false

$\{v_1, v_2, \boldsymbol{v_3}\}$ → true
$\{\boldsymbol{\bar{v}_1}, v_2, \boldsymbol{v_3}\}$ → true
$\{v_1, \boldsymbol{\bar{v}_2}, \boldsymbol{v_3}\}$ → true
$\{\boldsymbol{\bar{v}_1}, v_2, \bar{v}_3\}$ → true
$\{v_1, \boldsymbol{\bar{v}_2}, \bar{v}_3\}$ → true
$\{\boldsymbol{\bar{v}_1}, \boldsymbol{\bar{v}_2}, \bar{v}_3\}$ → true
$\{v_1, v_2, \boldsymbol{\bar{v}_4}\}$ → true
$\{\boldsymbol{\bar{v}_1}, \boldsymbol{\bar{v}_2}, \boldsymbol{\bar{v}_4}\}$ → true
$\{\boldsymbol{\bar{v}_2}, \boldsymbol{v_3}, v_4\}$ → true
$\{v_2, \bar{v}_3, v_4\}$ → false

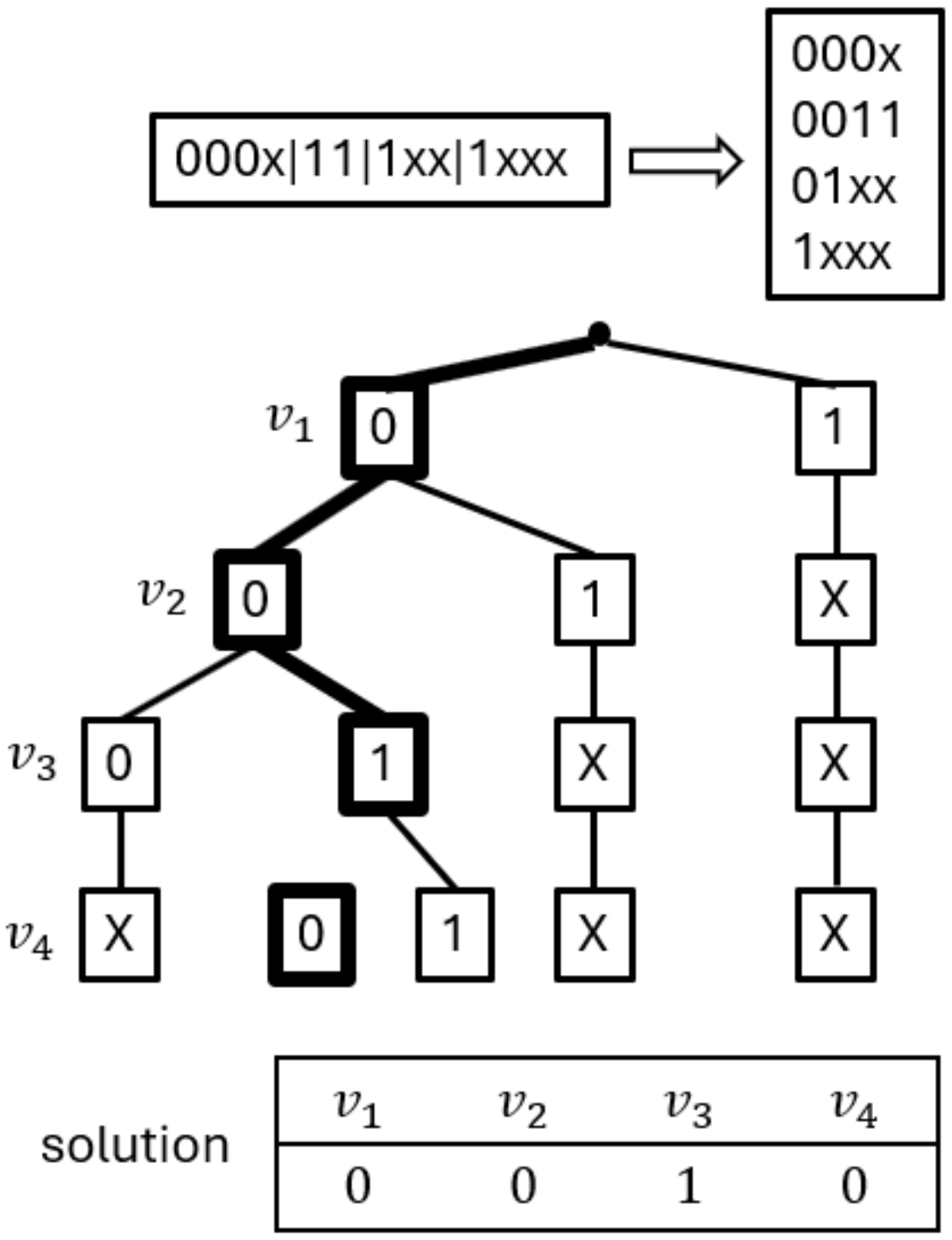

Figure 6 – Solution tree and a solution path.

## Binary Tree Size

It's important to notice that, as more and more clauses are added together, the total volume they cover is monotonically increasing, but the representation of that volume (in bits) tends to rapidly increase to some maximum and then decreases back down towards zero (assuming the problem has enough clauses and the final number of solutions is small).

We want the representation size to be as small as possible, and this depends on the processing order of the clauses. This is somewhat like a game of Tetris: we try to add clauses in a way that minimizes the total representation size, either by adding expressions that don't overlap much or by adding expressions that cause significant merging of branches.

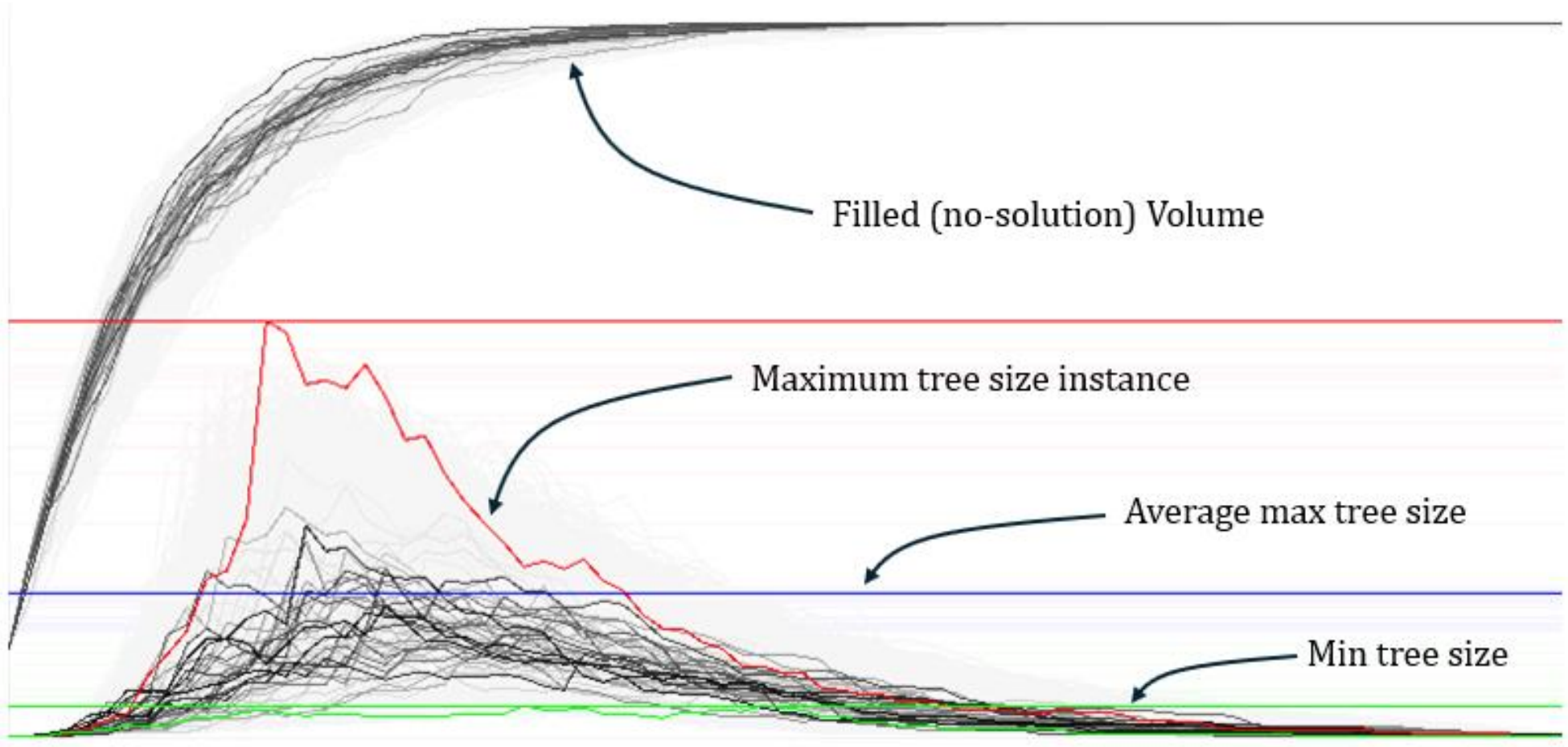

Figure 7 - Evolution of the solution tree size for the same problem instance but fordifferent random order of the clauses. In red, the largest tree size, in green the smallest, in blue the average size. Evolution of non-solution space volume is also shown.

Figure 7 shows the evolution of the tree size, for the same instance (N=16, M=80, two solutions), but with random shuffles of the order of the clauses. In red is the instance that leads to the largest transient representation. We also see the increase in filled volume (as a percentage), which starts at 0.125 with the first clause (since each clause occupies 1/8$^{th}$ of the total volume). The empty volume gives the solution count.

We see that the volume increase is pretty much independent of the clause order while the tree representation size typically will peak early on.

The steepest representation size increase will typically occur when adding clauses that do not have variables in common. The type of overlap induces a geometric growth at each clause (not considering some compression scheme that may be used in the representation).

For example for N = 12:

```
  000xxxxxxxxx
+ xxx000xxxxxx
_____________________________________________
  000xxxxxxxxx|1000xxxxxx|1x000xxxxxx|1xx000xxxxxx
+ xxxxxx000xxx
______________________________________________________________________
000xxxxxxxxx|1000xxxxxx|1000xxx|1x000xxx|1xx000xxx|1x000xxxxxx|1000xxx|1x000x
xx|1xx000xxx|1xx000xxxxxx|1000xxx|1x000xxx|1xx000xxx
+ xxxxxxxxx000
______________________________________________________________________
000xxxxxxxxx|1000xxxxxx|1000xxx|1000|1x000|1xx000|1x000xxx|1000|1x000|1xx000|
1xx000xxx|1000|1x000|1xx000|1x000xxxxxx|1000xxx|1000|1x000|1xx000|1x000xxx|10
00|1x000|1xx000|1xx000xxx|1000|1x000|1xx000|1xx000xxxxxx|1000xxx|1000|1x000|1
xx000|1x000xxx|1000|1x000|1xx000|1xx000xxx|1000|1x000|1xx000
```

This is why it pays to process clauses in an order that minimizes the subset of variables covered by those clauses. A smaller tree representation also means faster processing. Finding the right order of clause insertion is an optimization problem, with various potential heuristics.

One of them is to look at the set of clauses that still need to be processed, and pick the one that leads to the smallest increase of tree size. This seems like a decent strategy, but it costs

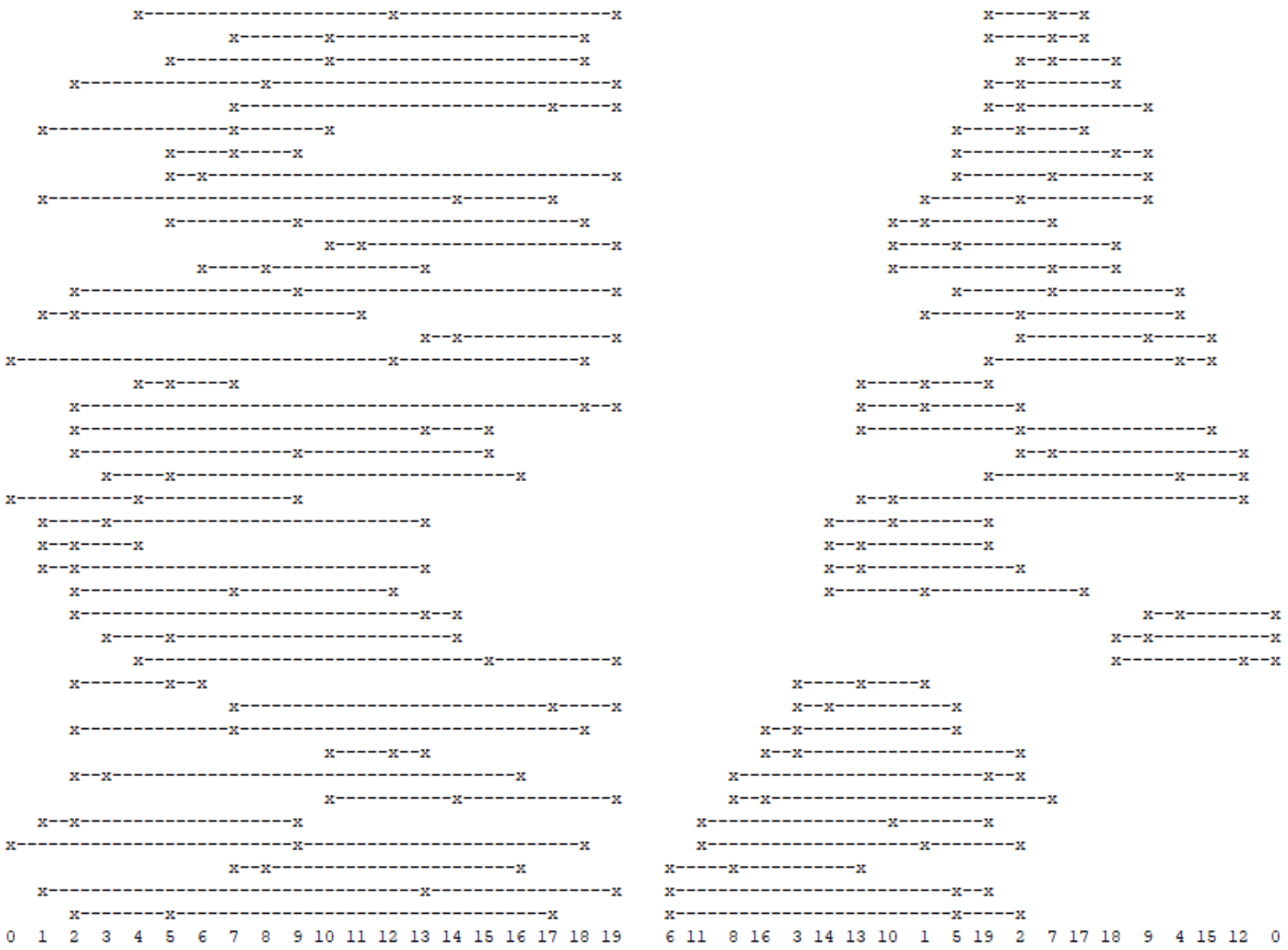

Figure 8 – set of variables and clauses before and after optimization. See procedure “*optimize_instance*” in the source code.

$O(M^2)$.
Another strategy consists in rearranging the variables and clauses based on some energy/relaxation incremental method in order to group clauses to increase local density, see Figure 8.
At any moment we only consider clauses that belong to a certain subset of variables, and then expand that subset by one. The subsets are considered based on clause “density”. Variables that are not involved in a set of clauses play no role in the growth of the tree, so it makes sense to process first clauses on the tightest variable subset and then grow that subset incrementally.

In Figure 9 we compare the tree size of the optimized clause/var order (blue) with the maximum (red) and minium (green) tree size found from random order of the clauses.

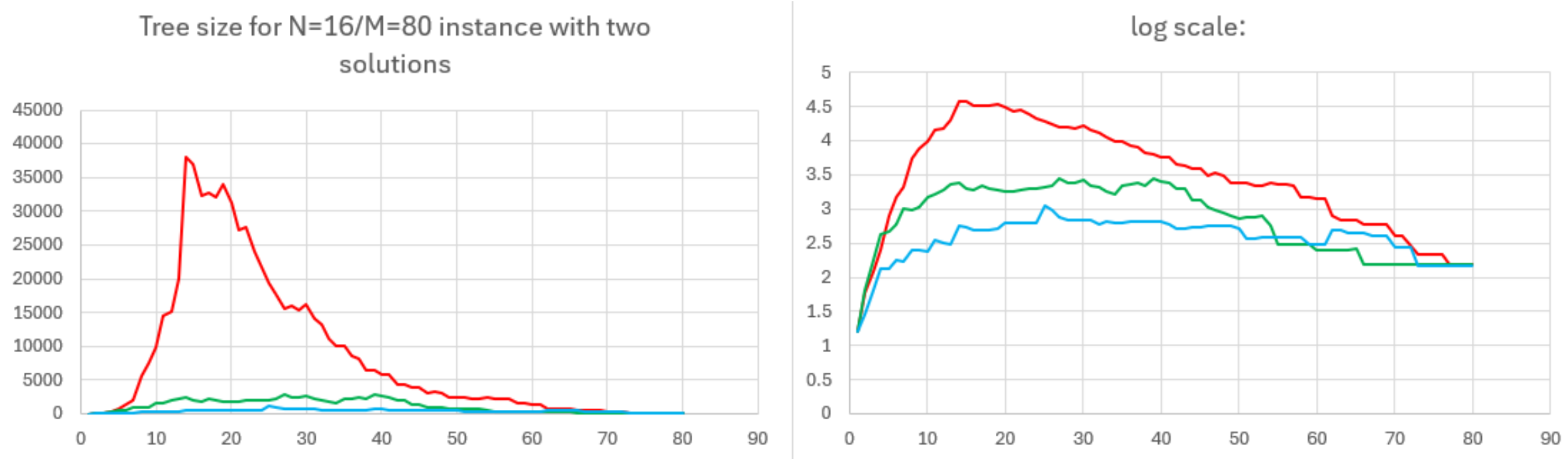

Figure 9 - For the same instance with16 variables and 80 clauses, in red and green the clause orders found randomly that lead to the largest and smallest tree size, in blue the optimized clause order.

In Figure 10 we illustrate the expected geometric growth of the average maximum tree size as a function of N, for optimized instances.

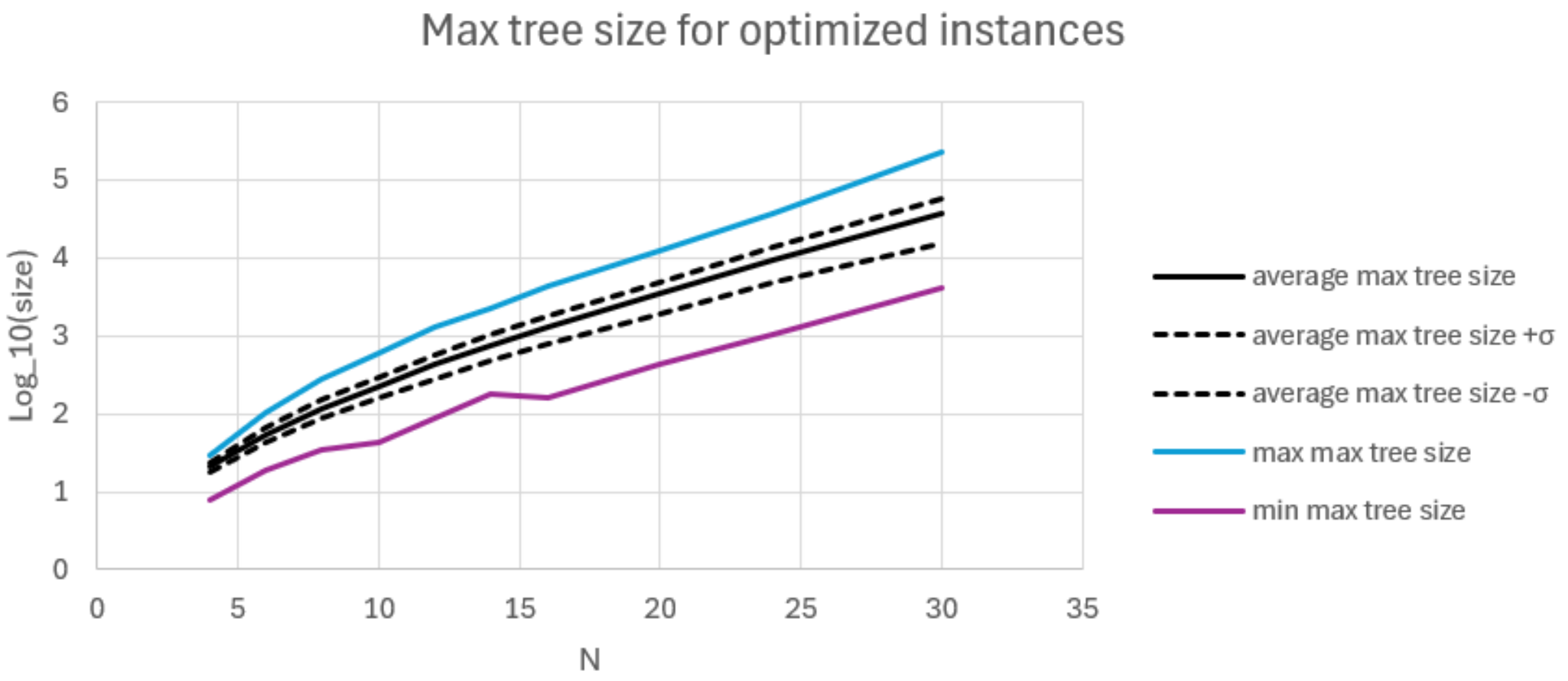

Figure 10 Evolution of average maximum tree size. Statistics for N are computed from 5000 random instances with optimized clause and variable ordering.

## Divide and (Hardly) Conquer:

The total volume considered ($2^N$) can be arbitrarily contracted by fixing some of the variables to 0 and 1. This scheme allows to deal with a smaller memory footprint for the binary trees. For each of those subspaces, we split the set of clauses.
In the example of Figure 11, we fixed variable v3.

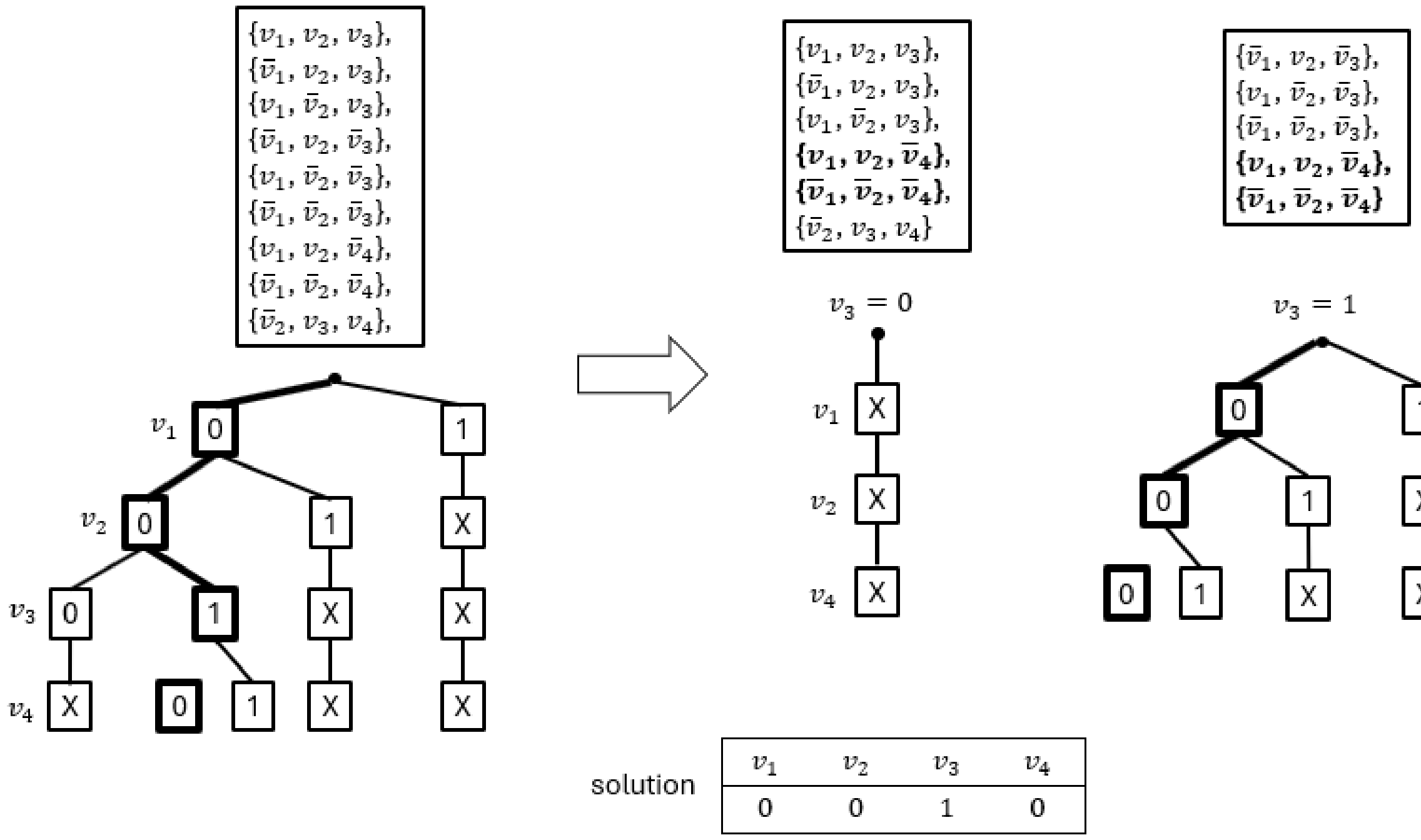

Figure 11 - Dividing the space in half by fixing a variable.

Such a split does not give any performance boost because, in general, when splitting along variable $v_i$, a significant number of clauses will be included both in the subset $v_i$ = 0 and $v_i$ = 1 (those clauses which have $v_i$ = x).
At the limit, if we fix all variables, each tree is reduced to a single leaf of volume 1 and will involve most of the clauses, and this is equivalent to doing backtracking (i.e. try all the permutations of variables and test on every clause).

## Overlaps and Counting Solutions

We saw previously that one can count the number of solutions by adding up the volume of each expression, like below

```
000x  → 1*1*1*2  = 2
0011  → 1*1*1*1  = 1
01xx  → 1*1*2*2  = 4
1xxx  → 1*2*2*2  = 8
```

occupied volume = 15
solution count = $2^4 - 15 = 1$

But this only works because those expressions do not overlap, i.e. all the clauses have at least one variable at a different value (0 or 1) than all the others.
Another approach consists in considering all the overlaps between clauses as negative volumes, and you need to also consider overlap of overlaps (as positive), etc.

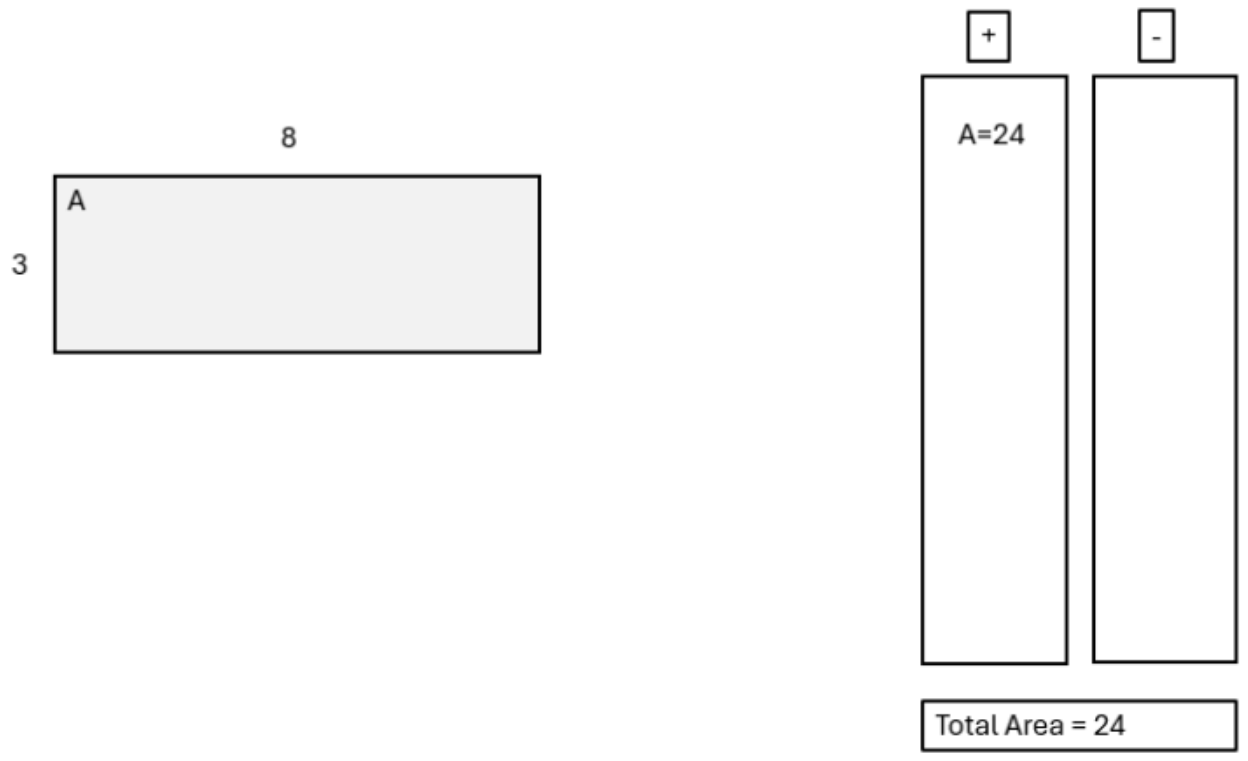

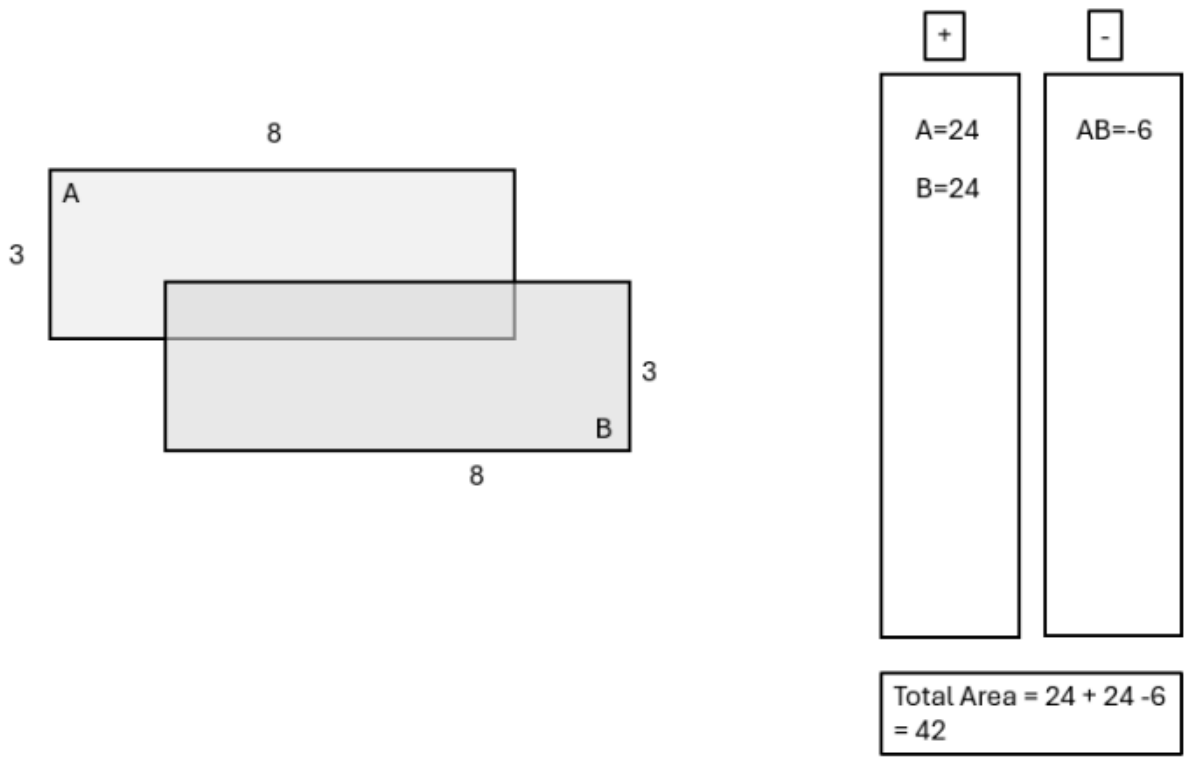

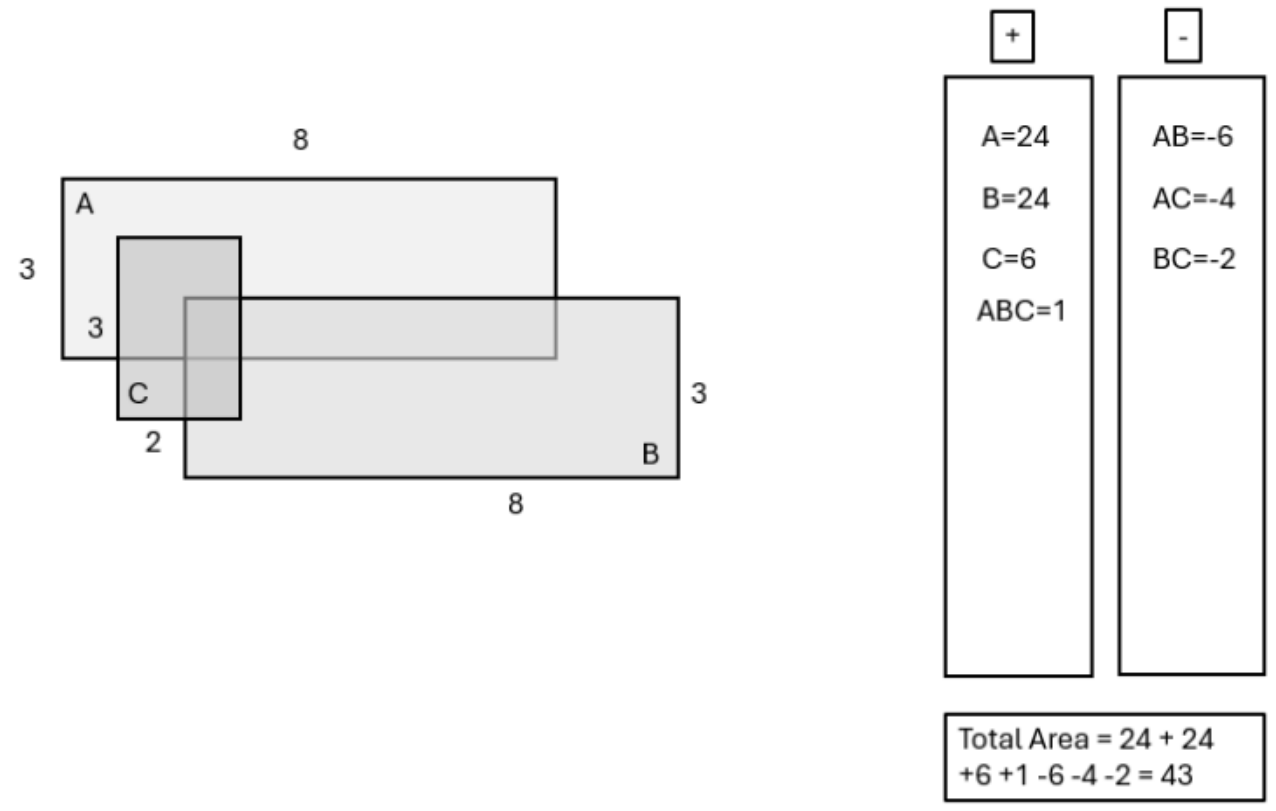

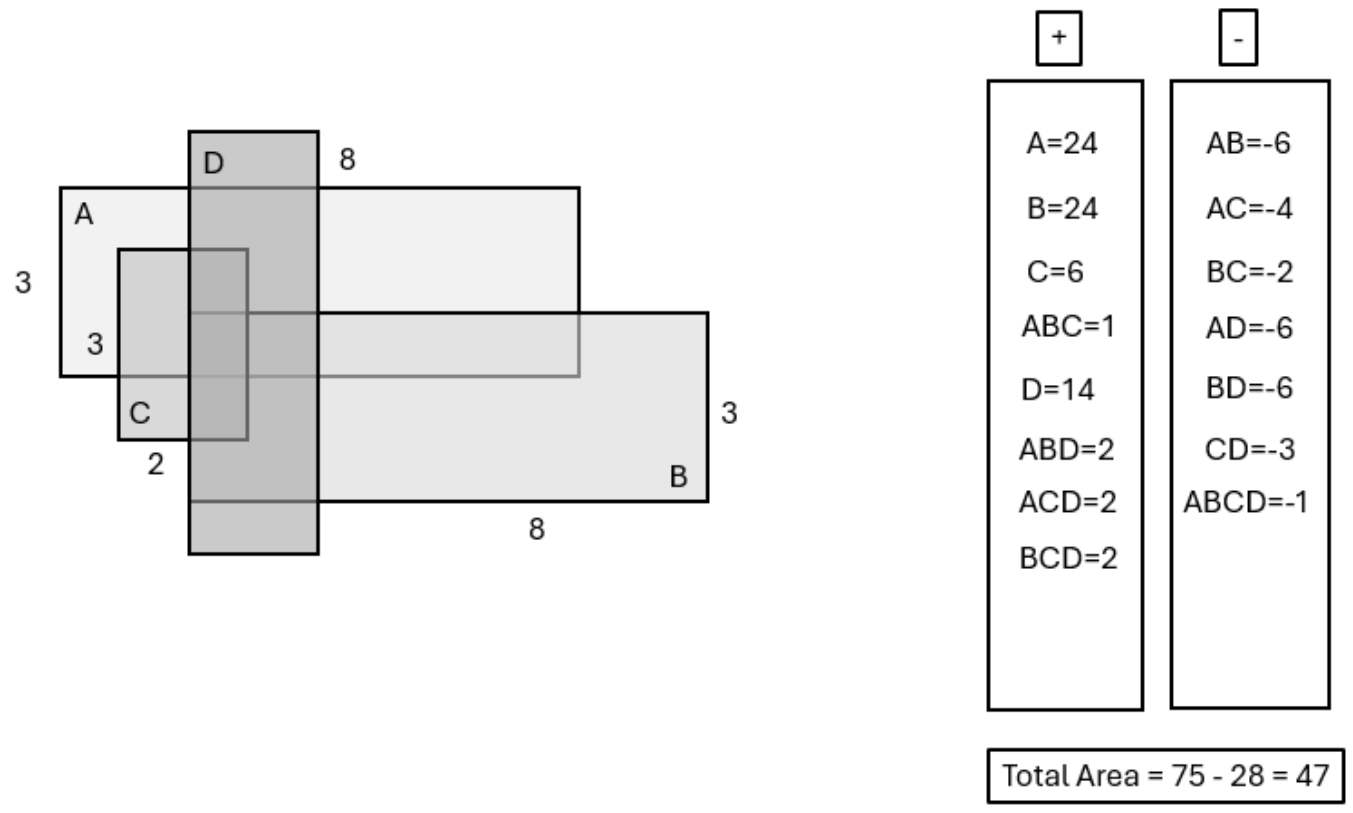

We can detect that some new surface is entirely contained in others, and drop it.

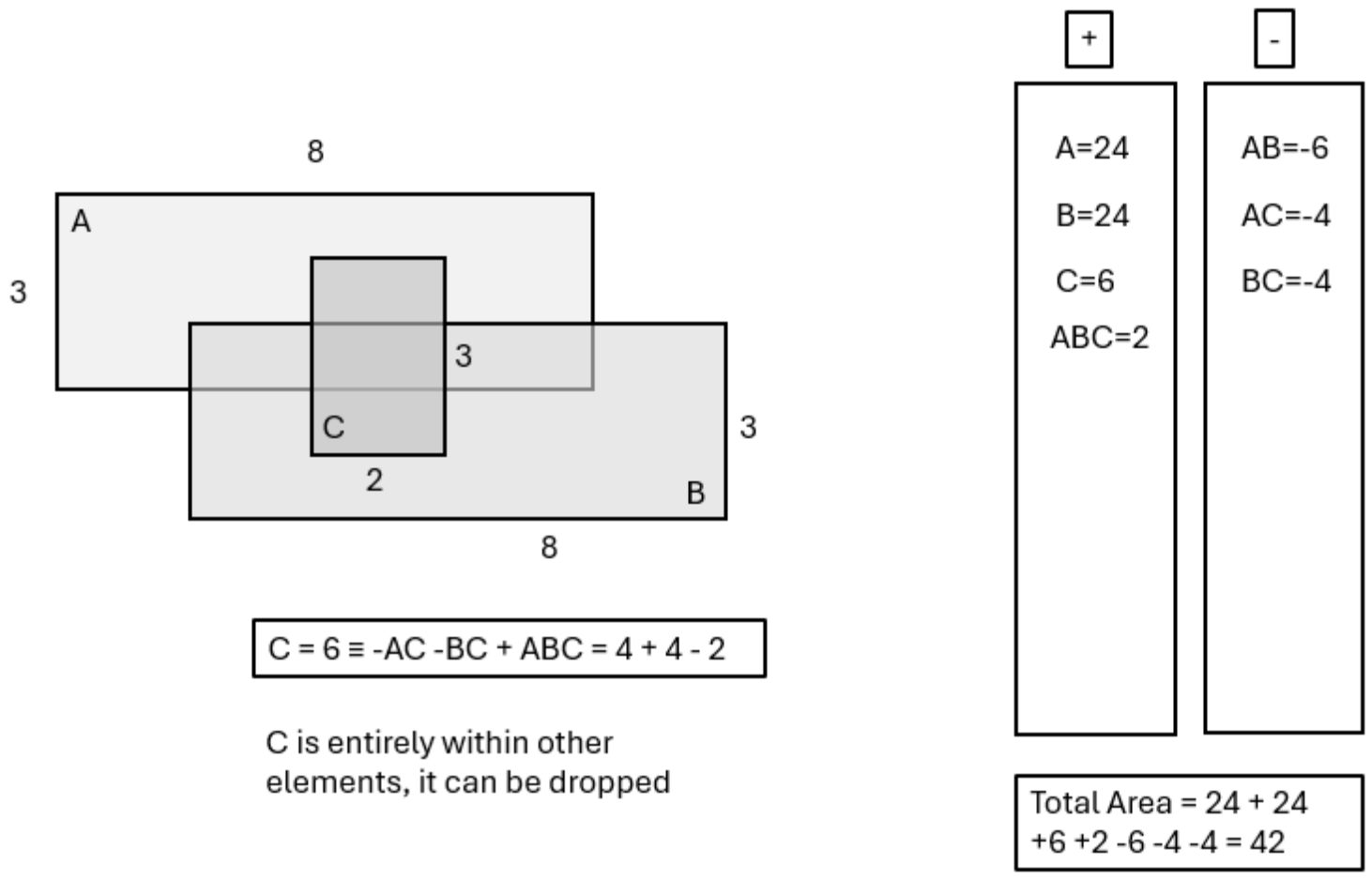

The same can be done with our expressions.

To generate the intersection of two expressions, we just repeat the variables at 0/1 that they have in common. E.g. intersection of '`x001xxx`' and '`00xx11x`' is '`000111x`'.
Our set of expressions leading to just one solution are

```
000x, 100x, 010x, 101x, 011x, 111x, 00x1, 11x1, x100
```

We add them one by one to the pair of lists (one positive and one negative)

```
+: 000x
-:

+: 000x, 100x
-:

+: 000x, 100x, 010x
-:

+: 000x, 100x, 010x, 101x
-:

+: 000x, 100x, 010x, 101x, 011x
-:

+: 000x, 100x, 010x, 101x, 011x, 111x
-:

+: 000x, 100x, 010x, 101x, 011x, 111x, 00x1
-: 0001

+: 000x, 100x, 010x, 101x, 011x, 111x, 00x1, 11x1
-: 0001, 1111

+: 000x, 100x, 010x, 101x, 011x, 111x, 00x1, 11x1, x100
-: 0001, 1111, 0100
```

The corresponding volumes are (each 'x' counts as a factor 2)

```
+: 000x, 100x, 010x, 101x, 011x, 111x, 00x1, 11x1, x100
   2     2     2     2     2     2     2     2     2      => tot = 18
-: 0001, 1111, 0100
   -1    -1    -1                                         => tot = -3
```

Total no-solution volume = 18 – 3 = 15
Solution count = $2^4$ – 15 = 1

One downside is that this method generates a lot of extra expressions (at geometric rate), but there is no need to sort. It's possible to mitigate this by merging expressions (this typically requires hashing the expressions).

## SAT/UNSAT Phase Transition, Solution Profile.

From a statistical point of view (if clauses are generated randomly), since the no-solution volume of the total space increases monotically with the number of clauses, if the number of clauses is quite low, the solution space will be very sparse and a lot of solutions exist. On the other hand, if the number of clauses is very high, it will be more likely that no solution exists.

It's been shown [1] that there is a sharp inflexion point where the statistics shifts from "most instances have solutions" (SAT) to "most instances have no solution" (UNSAT). This has been described as a phase transition, with an invariant (of N) for a clause count $M \sim 4.2$ N.

For a given N, we use a range of M, and randomly generate a large number of instances with M clauses, and look at the proportion of of instances that have solutions (rather than no solution), and plot the result.

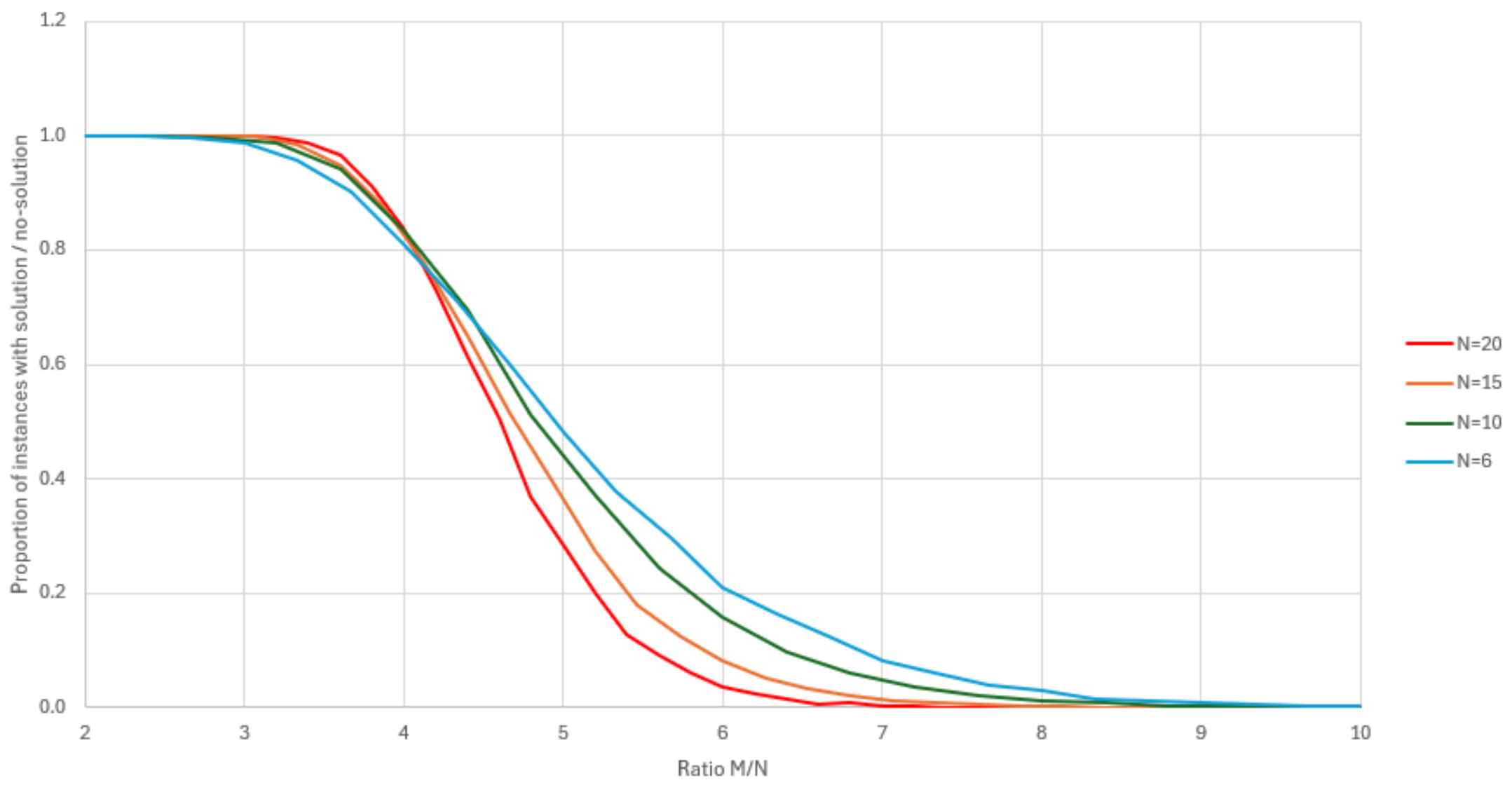

Figure 12 - Solution profile for various N and M, obtrained from solving random 3-SAT instances. See procedure "*compute_transition_stats*" in the source code.

## Simple Tiling Model

Using the geometric interpretation of 3-SAT, we can use an approximation that will reproduce its solution profile qualitatively (i.e. curves are the same shapes but exhibit a different transition point).

The model adopted considers a generic space of volume $2^N$, and we repeatedly add unit tiles randomly to try and fill the space. It's basically a "flattening" of the N dimension binary space.

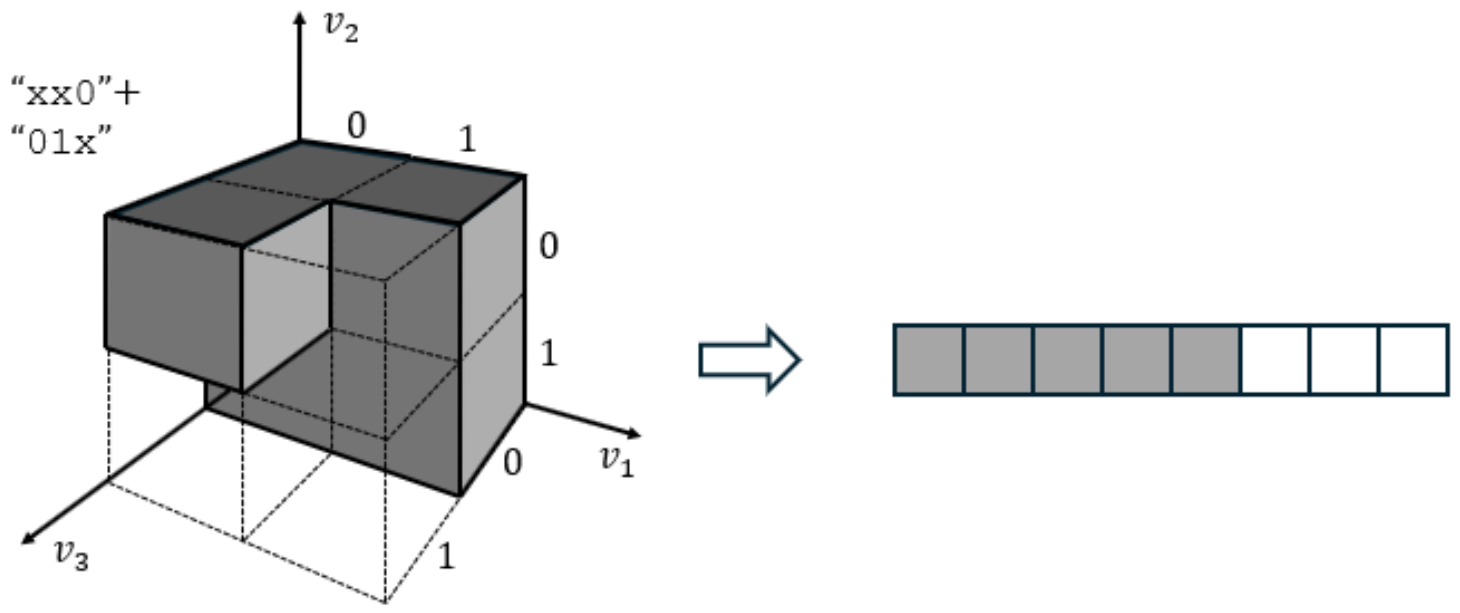

Figure 13 - The simple model flattens to solution space into a single dimension.

If the total volume is Vtot and the no-solution volume is V0, when we add an extra tile randomly, the probability of increasing the volume is p = (Vtot – V0)/Vtot, and the probability of overlap with the already filled volume is p = V0/Vtot.

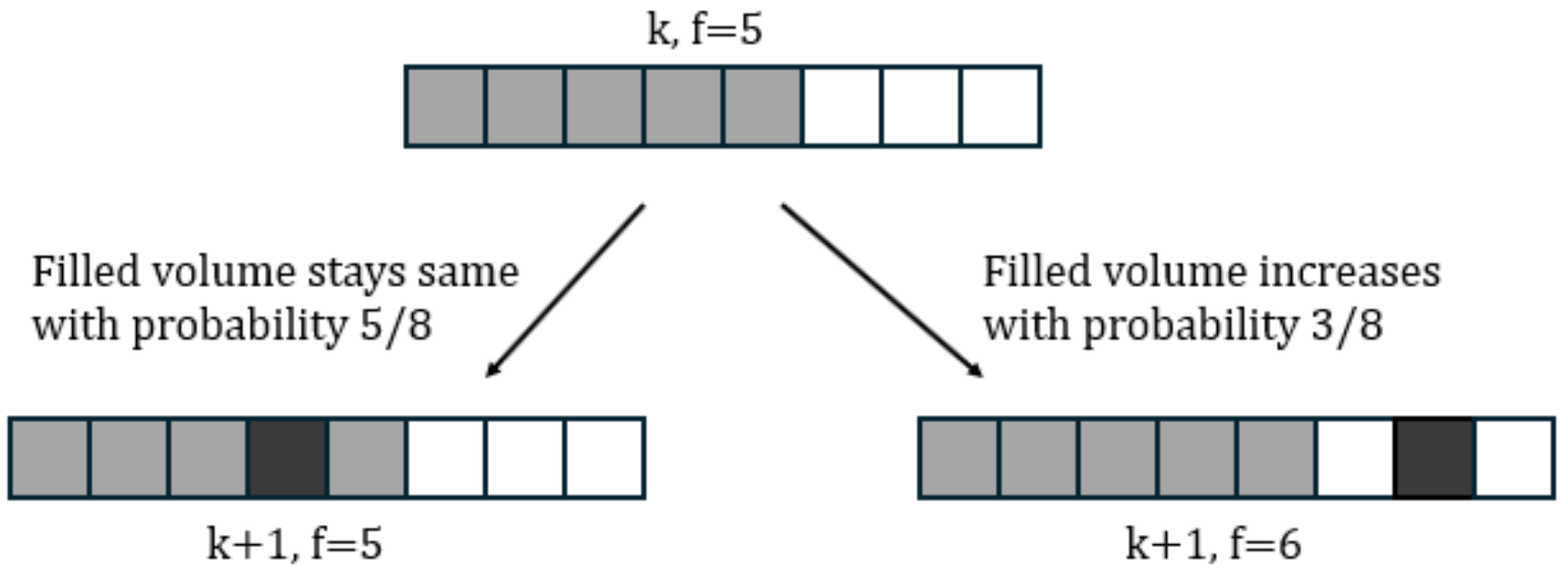

Figure 14 - Probability transitions for tile count = k and filled space f = 5.

For t = tile count, and f = filled volume, the probabilities evolve as

$$P(t+1, f+1) = P(t, f) * \left(1 - \frac{f}{2^N}\right) + P(t, f+1) * \left(\frac{f+1}{2^N}\right)$$

Giving the probability lattice of Figure 15.

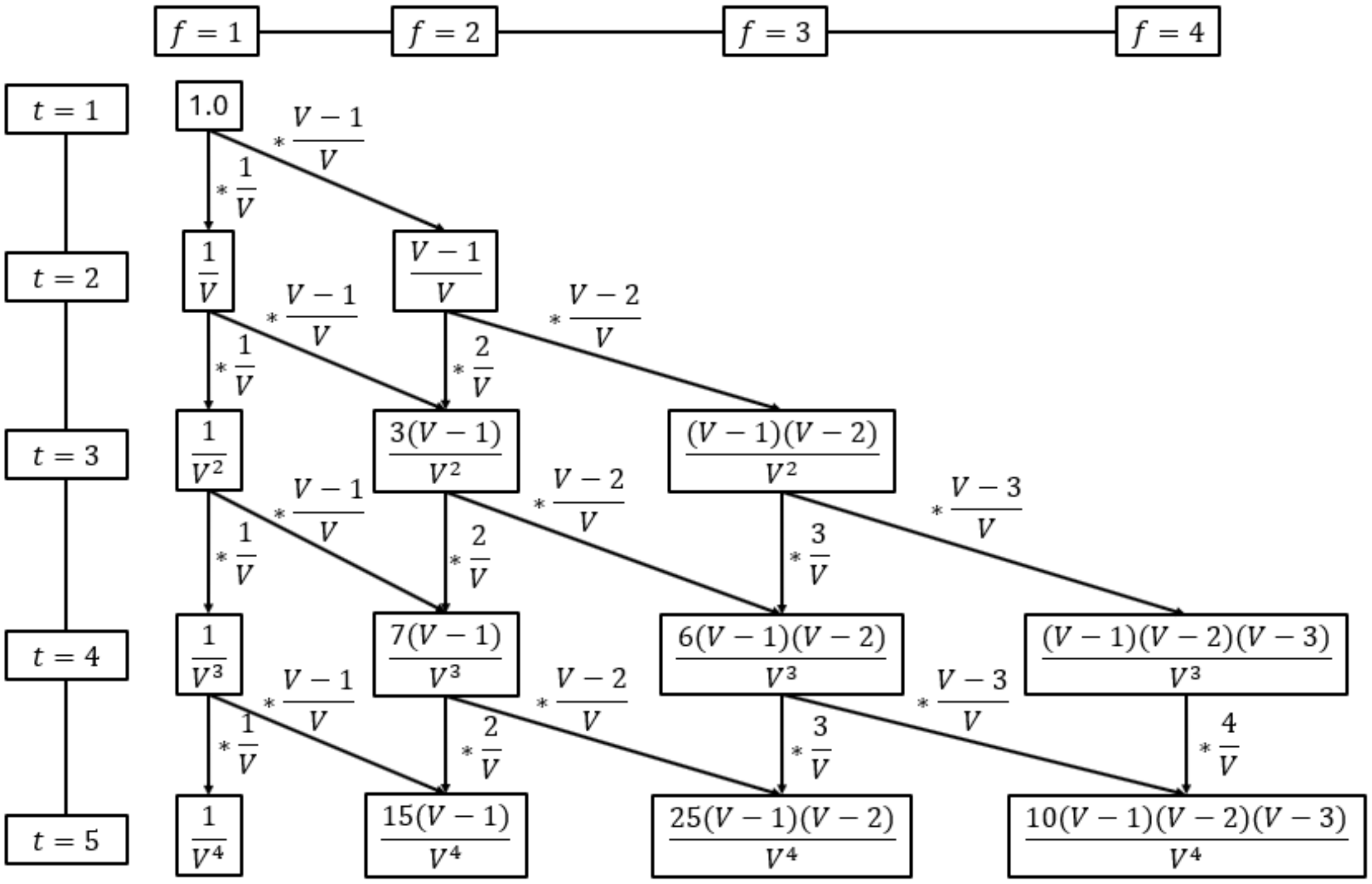

Figure 15 - Transition probability matrix for the simple tiling model.

The coefficients of that lattice are the "Stirling numbers of the second kind", S(t,f), which is the number of ways to partition t objects into f non-empty subsets.
The recurrence relation is S(t,f) = S(t-1,f-1) + f * S(t-1,f)
For example, S(4,3)=6, so there are 6 ways to partition 4 objects (a, b, c, d) into three different non-empty sets: a|b|cd, a|bd|c, ad|b|c, ab|c|d, ac|b|d, a|bc|d

| $S(t,f)$ | $f=1$ | $f=2$ | $f=3$ | $f=4$ | $f=5$ | $f=6$ | $f=7$ | $f=8$ | $f=9$ | $f=10$ |
|---|---|---|---|---|---|---|---|---|---|---|
| $t=1$ | 1 | | | | | | | | | |
| $t=2$ | 1 | 1 | | | | | | | | |
| $t=3$ | 1 | 3 | 1 | | | | | | | |
| $t=4$ | 1 | 7 | 6 | 1 | | | | | | |
| $t=5$ | 1 | 15 | 25 | 10 | 1 | | | | | |
| $t=6$ | 1 | 31 | 90 | 65 | 15 | 1 | | | | |
| $t=7$ | 1 | 63 | 301 | 350 | 140 | 21 | 1 | | | |
| $t=8$ | 1 | 127 | 966 | 1701 | 1050 | 266 | 28 | 1 | | |
| $t=9$ | 1 | 255 | 3025 | 7770 | 6951 | 2646 | 462 | 36 | 1 | |
| $t=10$ | 1 | 511 | 9330 | 34105 | 42525 | 22827 | 5880 | 750 | 45 | 1 |

The distribution we're looking for is given by the probabilities of the last column ($f = 2^N$), with $D(N,t) = 1 - p(t, 2^N)$
By writing explicitly the expression for the probabilities, we can derive that

$$D(N,t) = 1 - S(t, 2^N) * \frac{(2^N)!}{2^{tN}}$$

Where $S(t, 2^N)$ is the number of ways to group t objects in $2^N$ non-empty slots, $(2^N)!$ is the number of ways to arrange $2^N$ slots, and $2^{tN}$ is the total number of ways to assign t objects to $2^N$ slots, giving p(t, $2^N$) as the probability for t objects to fill a volume of size $2^N$.
Note that when t -> inf, D(N,t) -> 0, and we get the known assymptotic limit

$$S(t,f) \to \frac{f^t}{(f)!}$$

The relevant data is given by the last column in the probability lattice.

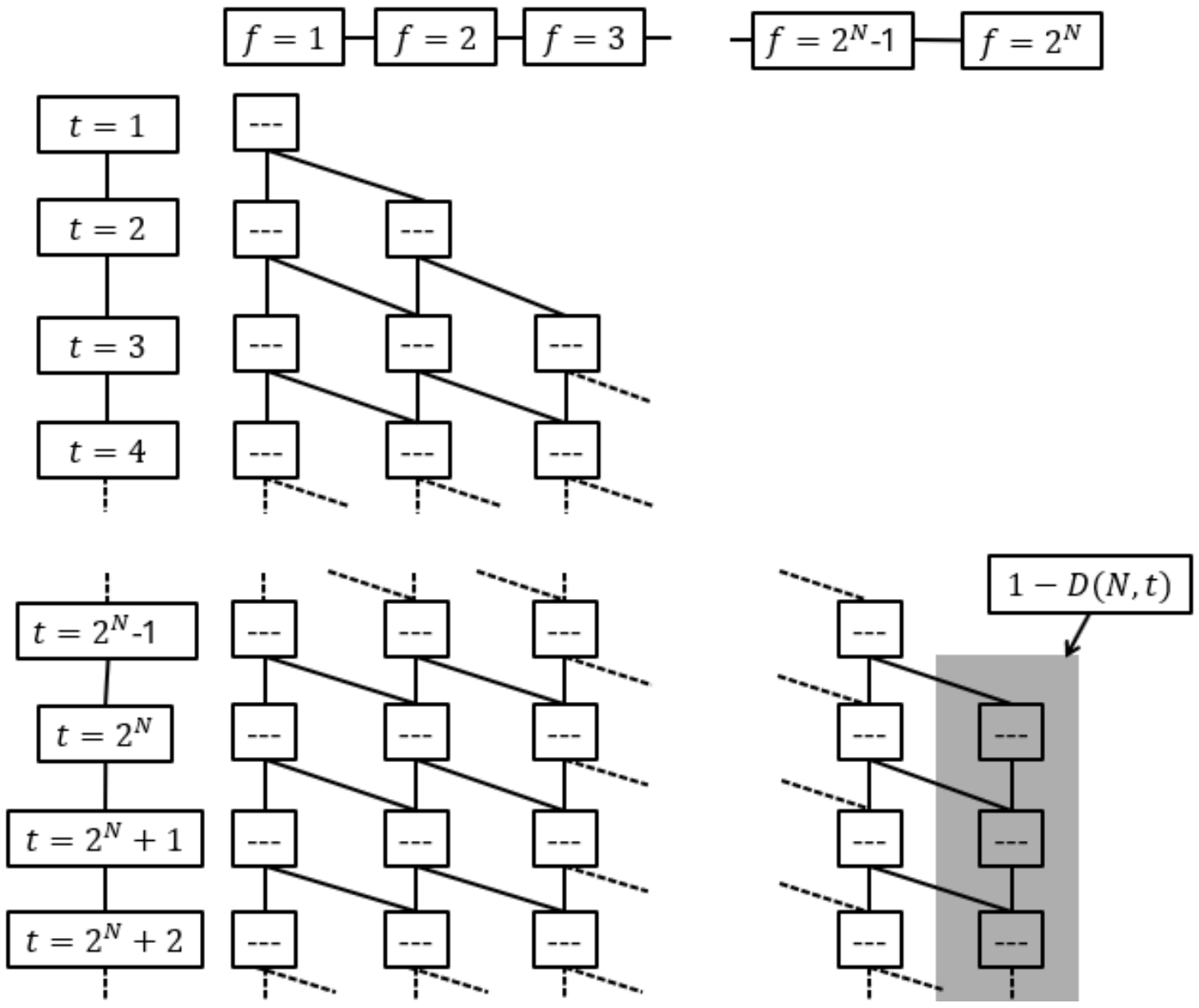

Figure 16 - Solution profile in the probability transition matrice.

We traced the curves for various values of N compared to the 3-SAT statistics

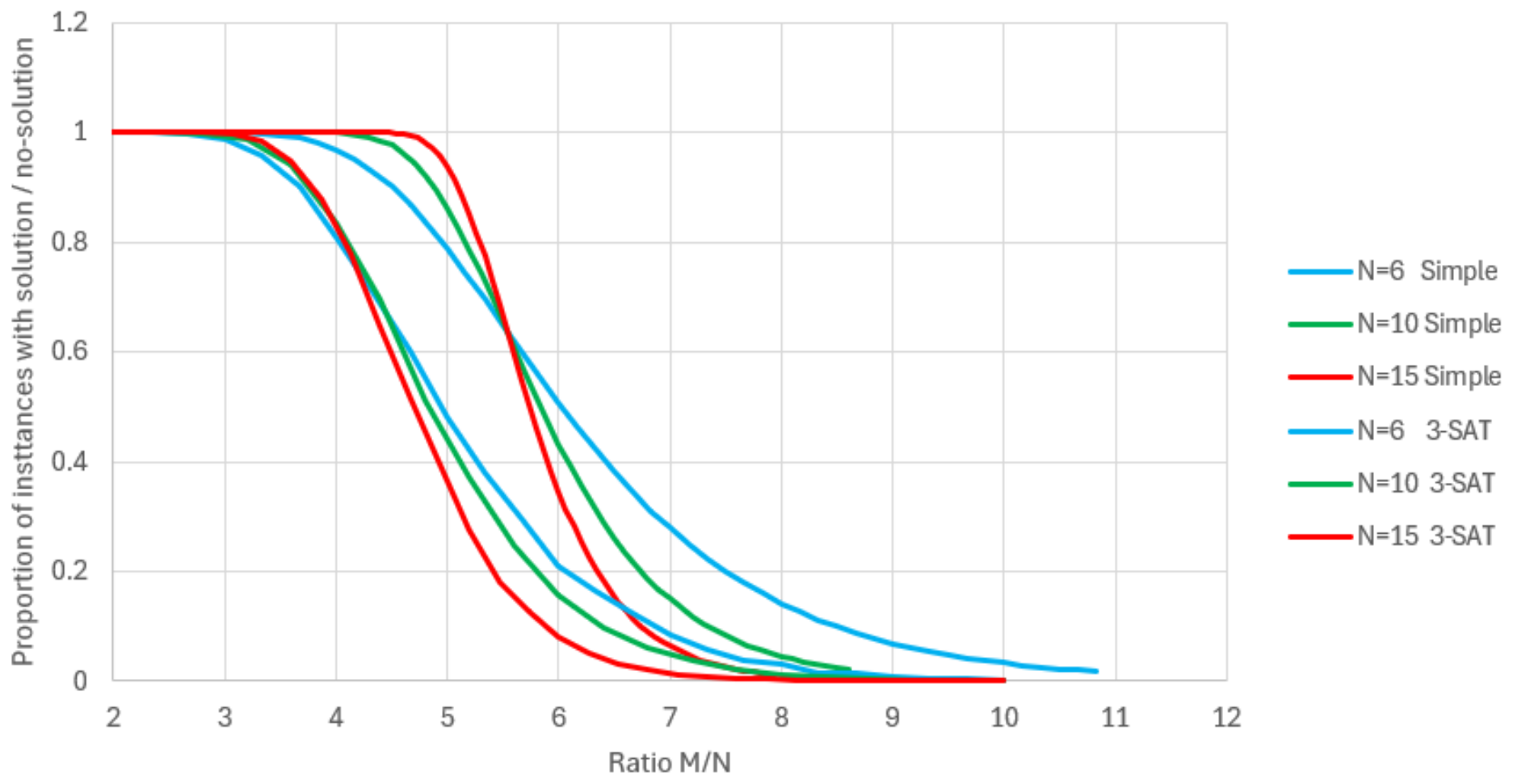

Figure 17 - Solution profile for 3-SAT and the simple tiling model. See procedure *"compute_3SAT_threshold_simple_approximation"* in the source code.

Note that this simple model does not internally consider clauses, and the invariant point is actually for a tile count $t' = \frac{t}{N2^N} \sim 0.6931$

The fact that we are considering triplet clauses introduces a constant factor $\frac{1}{2^3} = \frac{1}{8}$, giving a M/N = 0.693 x 8 = 5.545.

| $N$ | 2 | 3 | 4 | 5 | 16 |
|---|---|---|---|---|---|
| $Vt = 2^N$ | 4 | 8 | 16 | 32 | 65536 |
| $Vc = 2^{N-3}$ | 0.5 | 1 | 2 | 4 | 8192 |
| $D_0$ | 0.65 | 0.65 | 0.653 | 0.636 | 0.63212 |
| $t_0$ | 5.79 | 16.5 | 44 | 111 | 726768 |
| $M_0 = t_0/V_c$ | 11.6 | 16.5 | 22.5 | 27.7 | 89 |
| $M_0/N$ | 5.79 | 5.5 | 5.5 | 5.55 | 5.5451 |
| $t_0' = t_0/(N2^N)$ | 0.72 | 0.687 | 0.687 | 0.693 | 0.6931 |

The probability lattice for N=2 (which has no equivalent as a 3-SAT problem, but still exhibits the same characteristics):

| $t\backslash f$ | 1 | 2 | 3 | 4 | $D(N,t)$ |
|---|---|---|---|---|---|
| 1 | 1 | 0 | 0 | 0 | 1 |
| 2 | 0.25 | 0.75 | 0 | 0 | 1 |
| 3 | 0.0625 | 0.5625 | 0.375 | 0 | 1 |
| 4 | 0.015625 | 0.328125 | 0.5625 | 0.09375 | 0.90625 |
| 5 | 0.003906 | 0.175781 | 0.585938 | 0.234375 | 0.765625 |
| 6 | 0.000977 | 0.09082 | 0.527344 | 0.380859 | 0.619141 |
| 7 | 0.000244 | 0.046143 | 0.440918 | 0.512695 | 0.487305 |
| 8 | 6.1E-05 | 0.023254 | 0.35376 | 0.622925 | 0.377075 |
| 9 | 1.53E-05 | 0.011673 | 0.276947 | 0.711365 | 0.288635 |
| 10 | 3.81E-06 | 0.005848 | 0.213547 | 0.780602 | 0.219398 |
| 11 | 9.54E-07 | 0.002927 | 0.163084 | 0.833988 | 0.166012 |
| 12 | 2.38E-07 | 0.001464 | 0.123776 | 0.874759 | 0.125241 |
| 13 | 5.96E-08 | 0.000732 | 0.093564 | 0.905703 | 0.094297 |

The same data computed using the formula (N=2):

| $t'$ | $t$ | $St2(t,2^N)$ | $2^N!$ | $2^{tN}$ | $D(N,t)$ |
|---|---|---|---|---|---|
| 0.125 | 1 | 0 | 24 | 4 | 1 |
| 0.250 | 2 | 0 | 24 | 16 | 1 |
| 0.375 | 3 | 0 | 24 | 64 | 1 |
| 0.500 | 4 | 1 | 24 | 256 | 0.90625 |
| 0.625 | 5 | 10 | 24 | 1024 | 0.765625 |
| 0.750 | 6 | 65 | 24 | 4096 | 0.619141 |
| 0.875 | 7 | 350 | 24 | 16384 | 0.487305 |
| 1.000 | 8 | 1701 | 24 | 65536 | 0.377075 |
| 1.125 | 9 | 7770 | 24 | 262144 | 0.288635 |
| 1.250 | 10 | 34105 | 24 | 1048576 | 0.219398 |
| 1.375 | 11 | 145750 | 24 | 4194304 | 0.166012 |
| 1.500 | 12 | 611501 | 24 | 16777216 | 0.125241 |
| 1.625 | 13 | 2532530 | 24 | 67108864 | 0.094297 |

It is not obvious why there is an inflection point at Mo/N=Ko (or to/N2^N = ko, with K0=8*ko).
We can intuitively understand why the number of tiles needs to increase to fill the space, since the volume of a clause grows exponentially Vc=$2^{N-3}$ (each clause is 1/8 of the total space regardless of N). This gives the $2^N$ factor, but there's still the N factor.

There is a “dilution” effect as N increases by 1: to reach the same fill ratio probability, it’s just not enough to double the number of additional tiles.
E.g. for N=3, at half filled space, we have a 50% probability to go from 1/2 to 5/8 fill ratio.

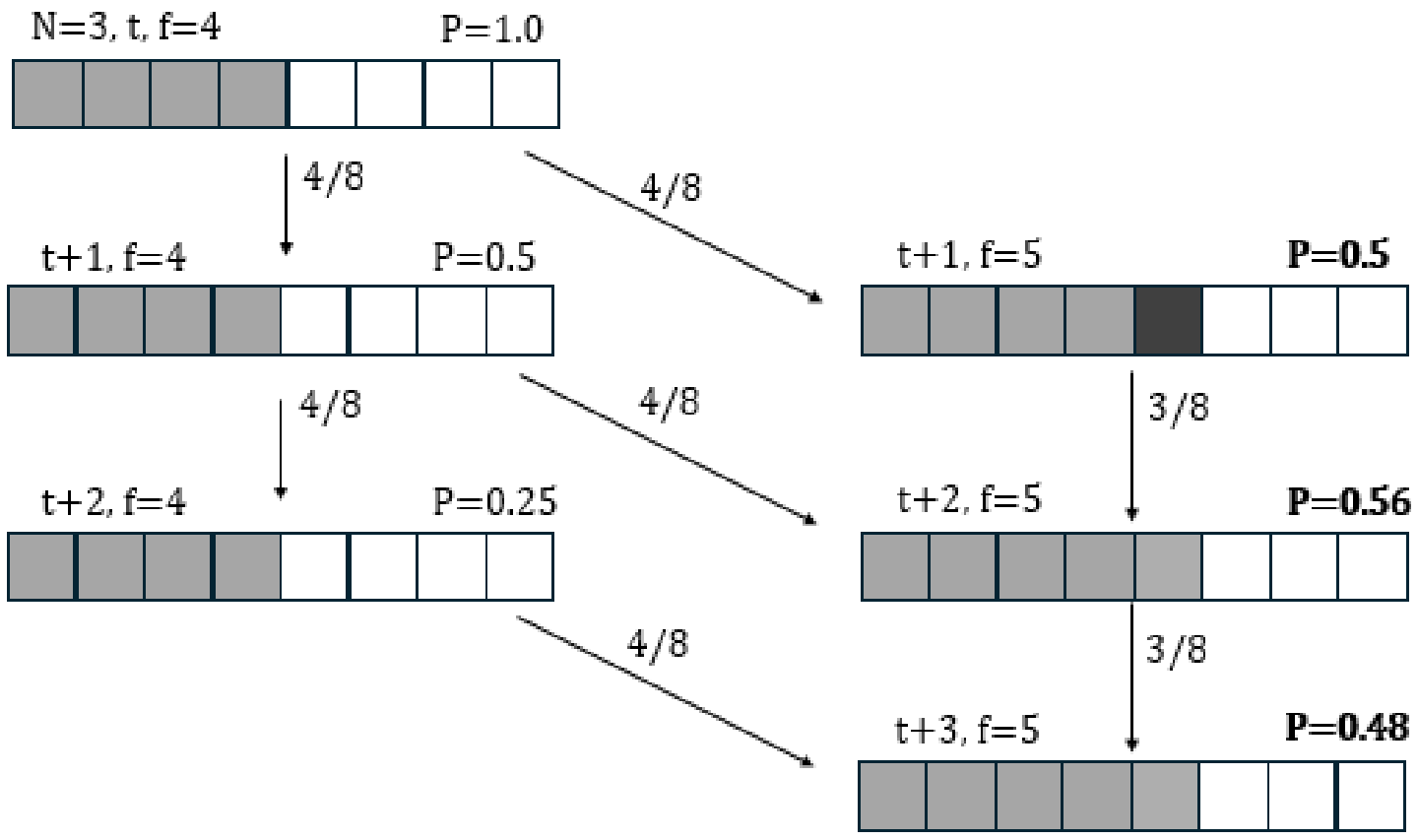

Figure 18 - Transitions for N=3

But when N=4 (we double the number of tiles), we can never reach 5/8=10/16 fill ratio with probability 50%

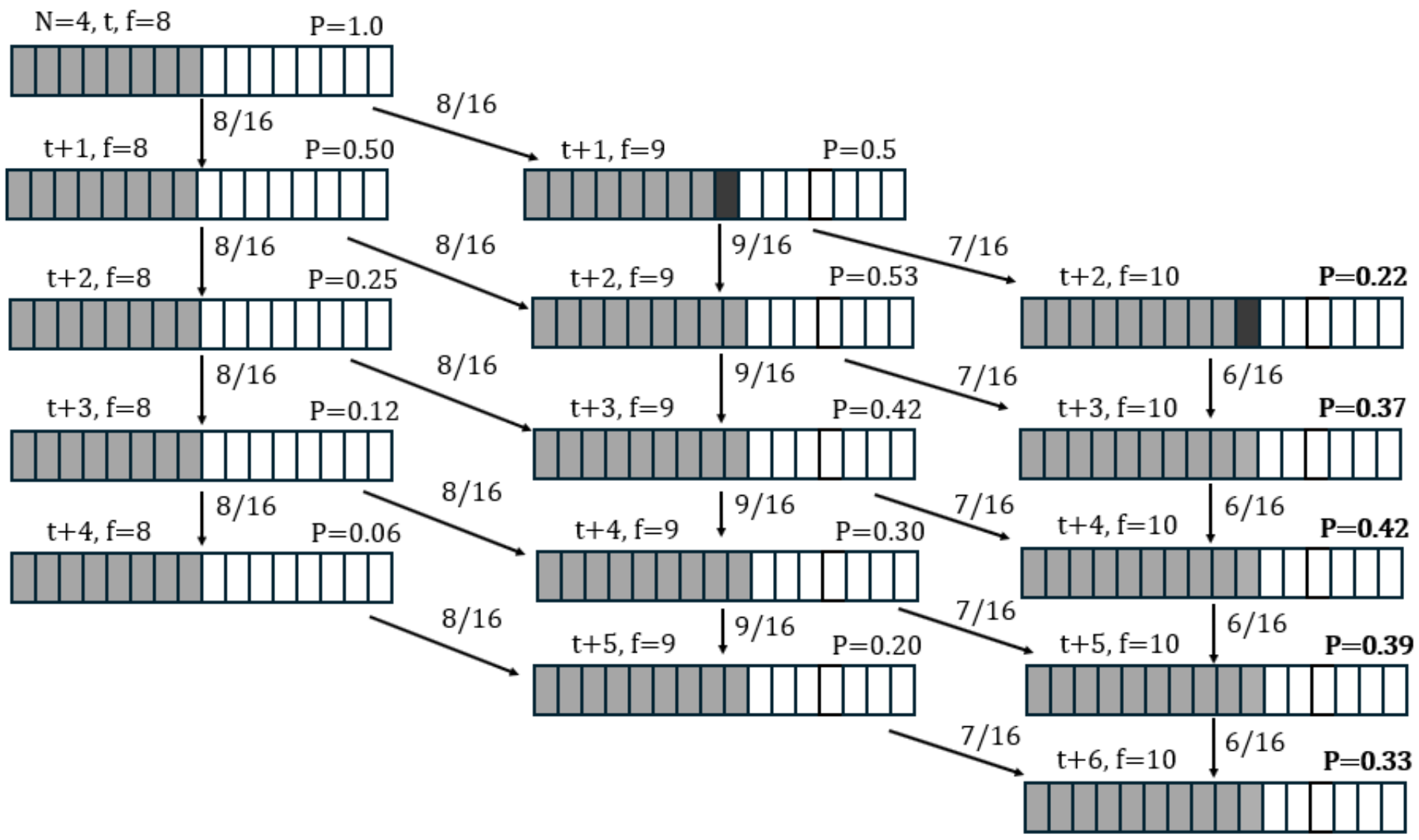

Figure 19 - Transitions for N = 4

Only on the last colum for $f = 2^N$, when the volume is full, the probabilities start to “compound” as t increases and will reach 1.0, asymptotically.

There is another probability lattice that gives the same D(N,t), with the recurrence relation

$$P(t+1, f+1) = P(t,f) * \left(\frac{f}{2^N}\right) + P(t, f+1) * \left(1 - \frac{f+1}{2^N}\right)$$

This lattice also leads to a different curve that meets D(t, N) at the same inflexion point (or very close to it):

$$M(N,t) = \frac{(2^N+1)^{t-1} - (2^N)! * S(t, 2^N+1)}{2^{Nt}}$$

Unlike D(N,t), it’s not known whether M(N,t) can be mapped to some measurable 3-SAT statistics.
One advantage is that the inflexion point can be estimated for the same value of N at the intersection of M(N,t) and D(N,t), at consecutive values of Stirling numbers.

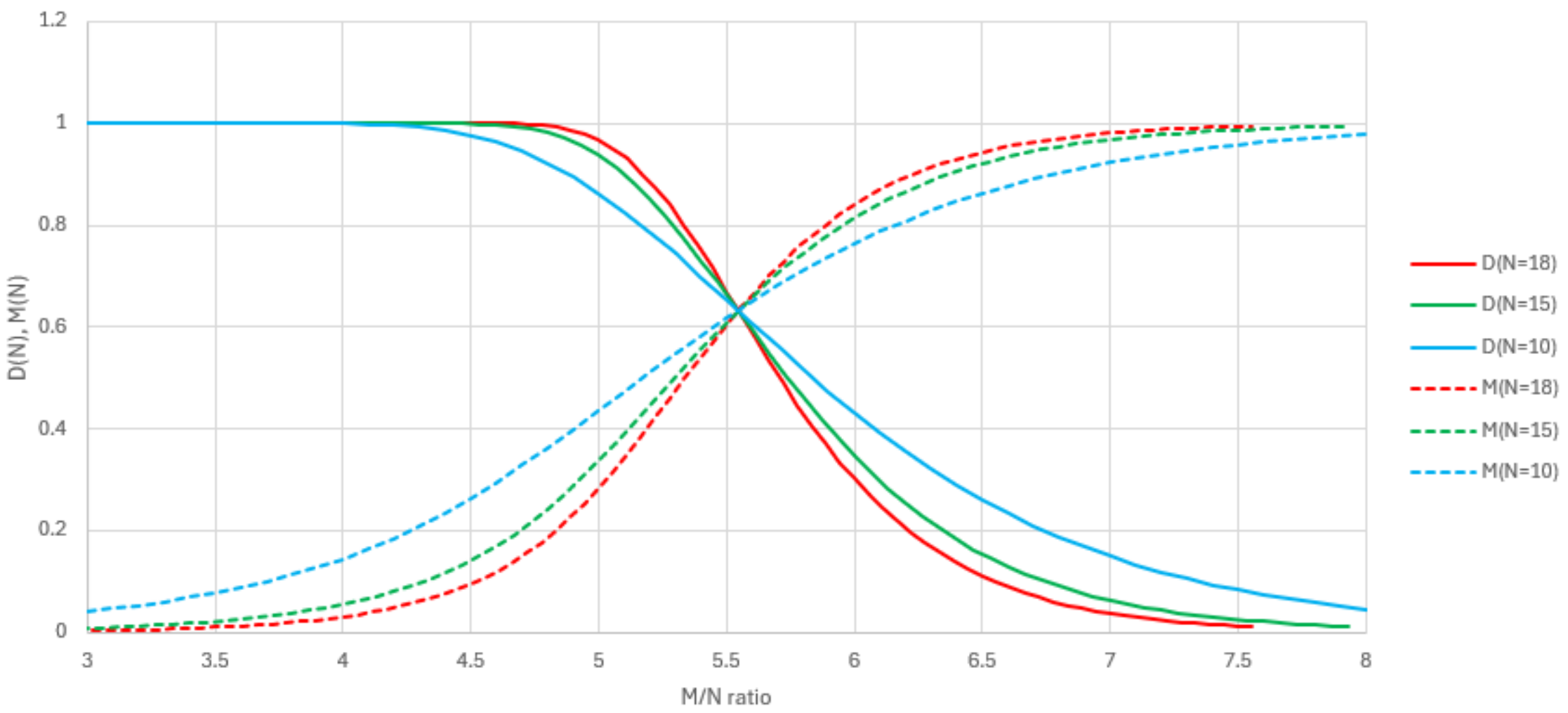

Figure 20 - D(N) and M(N) curves intersecting for the simple tiling model. See procedure “*compute_3SAT_threshold_simple_approximation_m_curve*” in the source code.

The transition point for this simple tiling model can be derived.
With

$$q = 2^N.$$

$$D(N,t) = 1 - \frac{q!\,S(t,q)}{q^t}$$

This has a straightforward probabilistic interpretation:

- Throw $t$ independent tiles uniformly into $q$ slots.
- $q!\,S(t,q)$ counts assignments in which every slot is occupied.
- Divide by $q^t$ to obtain the probability of filling every slot.

Thus $D$ is the probability that at least one slot remains empty.

There is a classical probability problem closely related to this model: the coupon collector problem.

For q slots, the characteristic number of random placements needed to fill all slots is approximately

$$t = q \ln q.$$

More precisely, introduce a scaling parameter c:

$$t = q \ln(q + c)$$

and consider the limit $t = q \to \infty$.
The probability that all slots have been occupied converges to

$$P_{full}(c) = \exp(-e^{-c})$$

Consequently,

$$D(c) = 1 - \exp(-e^{-c})$$

This already gives us a mathematically derived transition profile.

At c = 0,

$$D(0) = 1 - e^{-1}$$

and therefore

$$D(0) = 0.6321205588..$$

M itself involves (q+1) slots rather than q:

$$M(N,t) = \frac{(q+1)^{t-1} - q!\,S(t,q+1)}{q^t}$$

We can rewrite this using the probability that all (q+1) slots are occupied:

$$P_{full}^{(q+1)} = \frac{(q+1)!\,S(t,q+1)}{(q+1)^t}$$

Therefore,

$$M(N,t) = \frac{(q+1)^{t-1}}{q^t}\left(1 - P_{full}^{(q+1)}\right)$$

Now take the same asymptotic scaling:

$$t = q\ln(q+c)$$

There are two important limits.

First,

$$\frac{(q+1)^{t-1}}{q^t} \to \; e^c$$

Second, the coupon collector limit also applies to (q+1) slots:

$$P_{full}^{(q+1)} \to \; \exp(-e^{-c})$$

Hence,

$$M(c) = e^c(1 - \exp(-e^{-c}))$$

We now have explicit limiting expressions for both curves:

$$D(c) = 1 - \exp(-e^{-c})$$

$$M(c) = e^{c}(1 - \exp(-e^{-c}))$$

Set them equal:

$$1 - \exp(-e^{-c}) = e^{c}(1 - \exp(-e^{-c}))$$

For finite c, the common factor is strictly positive, so we can divide by it:

$$1 = e^{c}$$

Therefore

$$c = 0$$

For c = 0, t = $t_0$

$$t_0 = q \ln q$$

$$t_0 = 2^N \ln 2^N$$

$$t_0 = N\, 2^N \ln 2$$

$$\frac{t_0}{2\, 2^N} = \ln 2 = 0.6931471806 \ldots$$

The vertical coordinate follows from the expression for D:

$$D_0 = 1 - e^{-1} = 0.6321205588 \ldots$$

Finally, each 3-SAT clause covers $2^{N-3}$ tiles. Therefore the number of clauses corresponding to this tile count is

$$m_0 = \frac{t_0}{2^{N-3}}$$

$$\frac{m_0}{N} = 8 \ln 2 = 5.545177444 \ldots$$

Note that the intersection is not arbitrary. It occurs precisely when

$$t = q \ln q$$

which is the coupon collector's critical scale.

At this point, the expected number of empty slots is approximately

$$qe^{-t/q} = qe^{-\ln q} = 1$$

So, assymptotically, the intersection occurs when the expected number of unoccupied slots is exactly one.

## Less Simple Tiling Model

For the simplified model the invariant point is at (5.545, 0.632) instead of (4.2, 0.8) for 3-SAT.
The difference is due to the fact that 3-SAT has more spatial structure, i.e. each clause is a series of $2^{N-3}$tiles that are not spread randomly around in a flat 1D space but packed more regularly in a multidimentional space.
We can make our tiling model more realistic by making sure unit tiles belonging to the same clause do not overlap one another. This will modify the probability lattice and moves the inflexion point closer to the 3-SAT one.

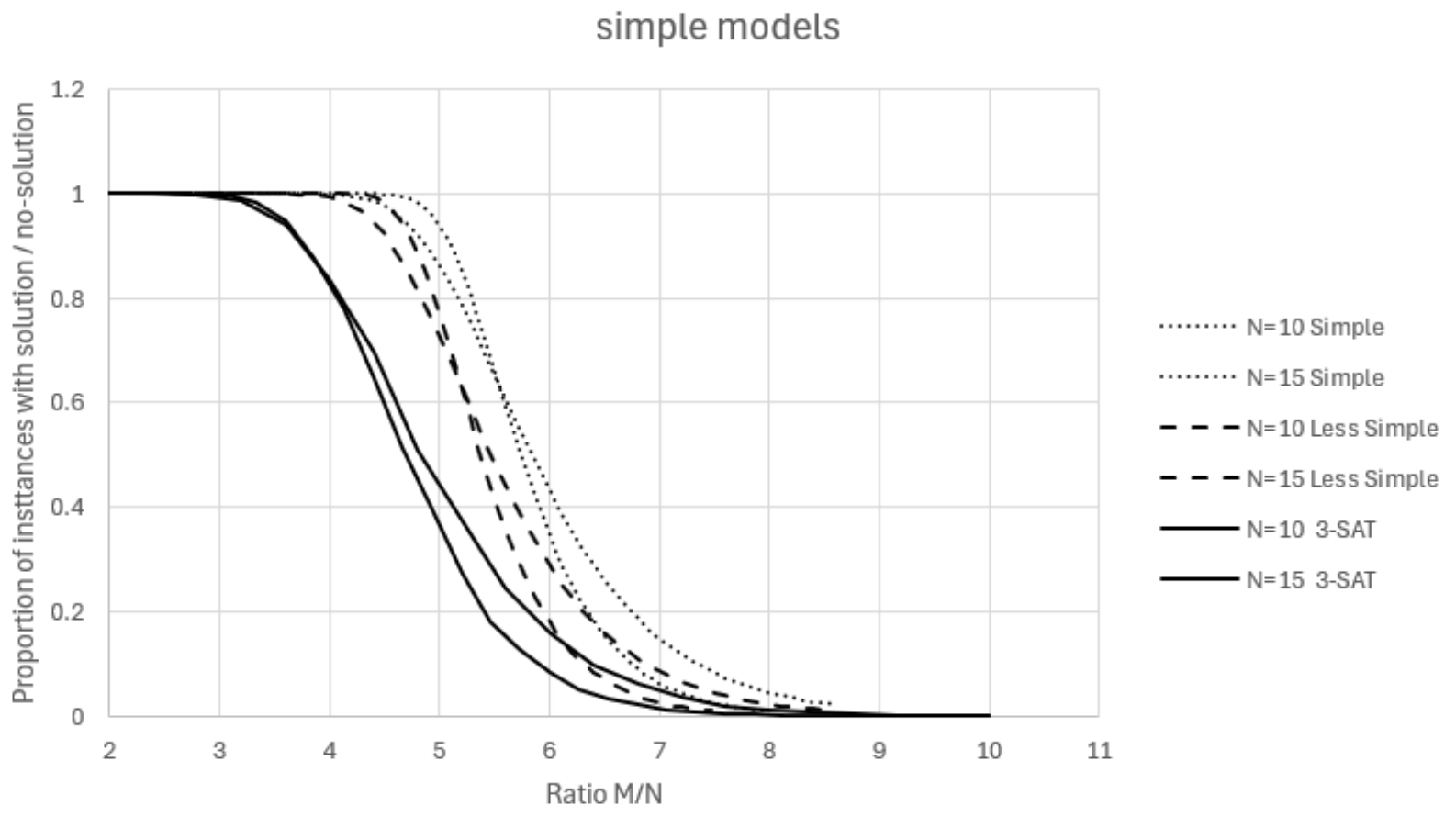

Figure 21 - Inflexion points for 3-SAT and our two simple tiling models. See
proc ce

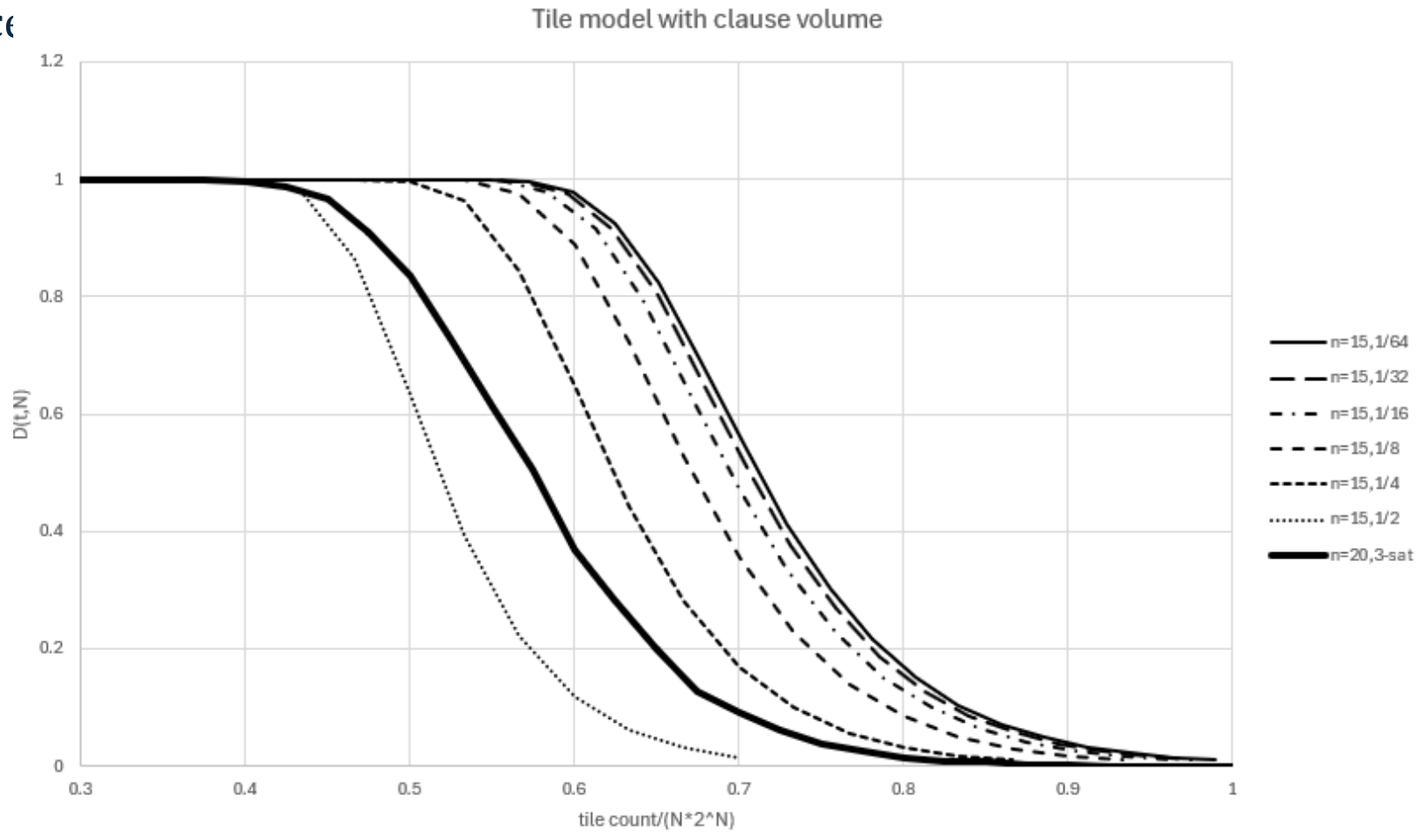

Figure 22 – Variants of tiling models with different element/clause sizes (from ½ of total volume to 1/64)

By modifying the clause/element volumes (e.g. for 3-sat this would be $1/8^{th}$ of the total volume), the D(N) curve can be shifted.

## Open Questions

- Theoretical maximum bounds for size of clause tree for 3-SAT.
- Full explanation of why the phase transition is at M/N = constant.